\documentclass[a4paper,11pt]{article}
\pdfoutput=1 

\usepackage{jheppub} 
\usepackage{bbm}
\usepackage[utf8]{inputenc}
\usepackage{xcolor}
\usepackage{cleveref}
\usepackage{subfigure}
\usepackage{braket}
\usepackage{hyperref}
\usepackage{graphicx}
\usepackage{caption}
\usepackage{tensor}
\usepackage{enumerate}
\usepackage{dsfont}
\usepackage{verbatim}
\usepackage{amsfonts,euscript,amssymb,stmaryrd,braket}

\usepackage{tikz,pgfplots}

\usetikzlibrary{decorations.pathreplacing,calligraphy}
\usetikzlibrary{decorations.markings}
\usepackage{resizegather}

\usepackage{slashed}
\usepackage{empheq}
\newcommand{\Tra}[1]{\mathrm{Tr}\left[#1\right]}

\def\Tr{\mathop{\mbox{\normalfont Tr}}\nolimits}
\definecolor{lightgreen}{HTML}{90EE90}

\newcommand{\Op}[1]{\mathcal{#1}}

\newcommand{\Aop}{\mathcal{A}}
\newcommand{\Top}{\mathcal{T}}

\newcommand{\unity}{\mathds{1}}

\newcommand{\dagg}{^\dagger}
\newcommand{\f}[2]{\frac{#1}{#2}}
\newcommand{\ave}[1]{\left<#1\right> }

\renewcommand{\d}{\text{d}}

\newcommand{\cbr}[1]{\left(#1\right)}

\newcommand\numberthis{\addtocounter{equation}{1}\tag{\theequation}}
\newcommand{\ra}{\rightarrow}

\renewcommand{\Re}[1]{\text{Re}\left\{#1\right\}}

\newcommand{\unm}{\frac{1}{2}}
\newcommand{\jacobitheta}[3]{\theta_{#1}\left(#2\left|#3\right.\right)	}
\renewcommand{\vec}[1]{{\boldsymbol{#1}}}

\renewcommand{\Tr}[1]{\text{Tr}\left(#1\right)}
\newcommand{\eva}[2]{\left.#1\right|_{#2}}
\usepackage{xcolor}
\hypersetup{
	colorlinks,
	linkcolor={blue!50!black},
	citecolor={blue!50!black},
	urlcolor={blue!80!black}
}
\newcommand{\Nop}{\Op{N}}
\usepackage{slashed}

\newcommand{\sgn}{\text{sgn}}

\usepackage{physics}
\hypersetup{colorlinks,bookmarksopen,bookmarksnumbered,
	citecolor=red,
	linkcolor=blue,
	pdfstartview=green,
	urlcolor=teal}

\begin{document}

	\title{\boldmath Entanglement resolution of free Dirac fermions on a torus}
	\vspace{.5cm}
	
	\author{Alessandro Foligno$^1$, Sara Murciano$^2$, and Pasquale Calabrese$^{3,4}$}
	\affiliation{$^{1}$School of Physics and Astronomy, University of Nottingham, Nottingham, NG7 2RD, UK}
	\affiliation{$^{2}$	Walter Burke Institute for Theoretical Physics, California
		Institute of Technology, Pasadena, California 91125, USA.}
	\affiliation{$^3$SISSA and INFN Sezione di Trieste, via Bonomea 265, 34136 Trieste, Italy.}
	\affiliation{$^{4}$International Centre for Theoretical Physics (ICTP), Strada Costiera 11, 34151 Trieste, Italy.}
	\vspace{.5cm}
	
	\abstract{Whenever a system possesses a conserved charge, the density matrix splits into eigenspaces associated to the each symmetry sector and we can access the entanglement entropy in a given subspace, known as symmetry resolved entanglement (SRE). 
		Here, we first evaluate the SRE for massless Dirac fermions in a system at	finite temperature and size, i.e. on a torus.
		Then we add a massive term to the Dirac action and we treat it as a perturbation of the massless	theory. The charge-dependent entropies turn out to be equally distributed among all the symmetry sectors at leading order. However, we find subleading corrections which  depend both on the mass and on the boundary conditions along the torus. 
		We also study the resolution of the fermionic negativity in terms of the charge imbalance between two subsystems. We show that also for this quantity, the presence of the mass alters the equipartition among the different imbalance sectors at subleading order. }
	
	1\maketitle
	\newpage
	\tableofcontents
	
	\section{Introduction}\label{Introduction}
	
	Entanglement is one of the most peculiar property of quantum physics. It is due to quantum correlations between different parts of a system, meaning that it is an intrinsically non-local quantity without a classical analogue. It is a fundamental tool in theoretical physics, with applications  in high-energy physics \cite{high1,high2,high3,high4}, in the context of black holes \cite{black1,black2,Solodukhin_2011}, and in low-energy physics, where it can be useful to study extended quantum systems \cite{intro1,intro2,eisert-2010,intro3,Vidal_2003}.
	
	Given the relevance of the subject, several different ways to quantify the entanglement have been studied over the years. Here, we consider two of them: the first one is the von Neumann entropy, which characterises the entanglement of a bipartite system in a pure quantum state $\ket{\psi}$ \cite{intro3}. Starting from a bipartition of a system $A\cup B$, and the corresponding reduced density matrix (RDM) $\rho_A=\mathrm{Tr}_B(\ket{\psi}\bra{\psi})$, the entanglement entropy is defined as
	\begin{equation}
		\label{eq:entr0}
		S_1=-\Tra{\rho_A\ln \rho_A}.
	\end{equation}
	A related family of functions, known as R\'enyi entropies, is given by
	\begin{equation}
		\label{eq:entr}
		S_n=\f{1}{1-n} \ln\mathrm{Tr} \rho_A^{n}, \quad S_1 =\lim_{n \to 1} S_n=-\Tr[\rho_A\log \rho_A].
	\end{equation}
	The latter are particularly convenient because in the path-integral formalism, for integer $n$, $\Tra{\rho_A^{n}}$ corresponds to a partition function on an $n$-sheeted Riemann surface $\mathcal{R}_n$, obtained by joining cyclically the $n$ sheets along the region $A$ \cite{Calabrese:2004eu}. In quantum field theory, this object can be computed as a correlation function involving a particular type of twist fields, which are related to branch points in the Riemann surface $\mathcal{R}_n$ \cite{Cardy:2007mb,Calabrese_2009}. Then, the limit $n \to 1$ allows to compute the von Neumann entropy.
	
	Despite the success of R\'enyi entropies in describing bipartite pure systems, entanglement entropies are no longer good measures of entanglement in mixed states since they mix quantum and classical correlations (e.g. in a high temperature state, $S_1$ gives the extensive result for the thermal entropy that has nothing to do with entanglement). 
	If we consider the mixed state described by the reduced density matrix $\rho_A$, where $A=A_1 \cup A_2$, and we want to quantify the entanglement between $A_1$ and $A_2$, we can use the entanglement negativity \cite{vw-02,Plenio_2005}, defined as
	\begin{equation}\label{eq:bos1}
		\mathcal{N}^{(b)}\equiv\f{\mathrm{Tr}|\rho_A^{T_1}|-1}{2},
	\end{equation}
	where $\mathrm{Tr}|O|:= \mathrm{Tr} \sqrt{O^{\dagger}O}$ denotes the trace norm and $\rho_A^{T_1}$ is the partially-transposed RDM of $A$, which is defined as follows. 
	Let us write the RDM as 
	\begin{equation}		\rho_A=\sum_{ijkl}\mel{e^1_i,e^2_j}{\rho_A}{e^1_k, e^2_l}\ketbra{e_i^1,e_j^2}{e^1_k,e^2_l},
	\end{equation}
	where $\ket{e_j^1}$ and $\ket{e_k^2}$ are orthonormal bases for the Hilbert spaces $\mathcal{H}_1$ and $\mathcal{H}_2$ describing the system in $A_1,A_2$, respectively. The partial transpose $\rho_A^{T_1}$ is defined by exchanging the matrix elements in the subsystem $A_1$, i.e.
	\begin{equation}\label{eq:bosonic}
		(\ket{e_i^1,e_j^2}\bra{e^1_k,e^2_l})^{T_1} \equiv \ket{e_k^1,e_j^2}\bra{e^1_i,e^2_l}.
	\end{equation}
	In terms of the eigenvalues of $\rho_A^{T_1}$, $\lambda_i$, and recalling the normalisation $\sum_i \lambda_i=1$, $\mathrm{Tr}|\rho_A^{T_1}|$ can be written as
	\begin{equation}
		\mathrm{Tr}|\rho_A^{T_1}|=\sum_i |\lambda_i|=\sum_{\lambda_i >0}|\lambda_i|+\sum_{\lambda_i <0}|\lambda_i|=1+2\sum_{\lambda_i<0}|\lambda_i|.
	\end{equation}
	This expression shows that the negativity measures ``how much'' the eigenvalues of $\rho_A^{T_1}$ are negative, from which the name negativity comes. 
	Finally, it is worth mentioning that the superscrit $b$ in Eq. \eqref{eq:bos1} refers to the fact that this expression is sometimes referred to as bosonic negativity: indeed, it turned out that its definition is not well-suited to investigate entanglement properties in the context of (free-)fermionic systems. To circumvent this issue, a slightly different definition for the partial transpose, based on the partial time-reversal (TR) transformation of the RDM, can be adopted, as we are going to review in Section \ref{sec:defs}.
	
	Recently, it has also gained importance the study of the interplay	between entanglement and the internal symmetries of a system. This new aspect has become accessible experimentally \cite{Lukin,neven2021symmetryresolved,vitale2021symmetryresolved,operator_entanglement}, together with the development of the theoretical framework necessary to treat the problem \cite{Goldstein_2018,PhysRevB.98.041106}. In this regard, in \cite{Goldstein_2018} a simple generalisation of the replica trick has been proposed to relate symmetry resolved quantities to the moments of 
	$\rho_A$ on a modified Riemann surface: we refer to them as {\it charged moments}. 
	This idea paved the way to the study of the {\it symmetry resolved entanglement} in various contexts, such as Conformal Field Theories (CFTs) \cite{Goldstein_2018,PhysRevB.98.041106,Belin:2013uta,Murciano_2020,Capizzi_2020,calabrese2021symmetryresolved,Milekhin2021,Hung:2019bnq,ghasemi}, interacting integrable quantum field theories 
	\cite{Horv_th_2020,horvath2021u1,Capizzi:2021kys,Horvath:2021rjd,cdfmssca-22,cdfmssca-22-2}, holographic settings \cite{Zhao_2021,Weisenberger:2021eby,Zhao:2022wnp,Baiguera:2022sao}, disordered systems \cite{Turkeshi_2020,Kiefer2_Emmanouilidis_2020,Kiefer-Emmanouilidis20Scipost,Kiefer-Emmanouilidis21}, non-trivial topological phases \cite{Cornfeld_2019,Monkman_2020,Azses_2020,ahn-20,ads-21}, spin chains \cite{Bonsignori_2019,Bonsignori:2020laa,Fraenkel_2020, Murcianobis_2020,Murciano:2020lqq,Calabresebis_2020,Laflorencie_2014,Barghathi_2018,Barghathi_2019,Barghathi_2020,PhysRevB.103.L041104,parez2021exact,Wiseman_2003,Estienne_2021,Tan_2020,fraenkel2021entanglement,ore-21,Ares:2022hdh,Scopa_2022,jones-22,pvcc-22,Lau2022,page,asymmetry}, and also its extension to several entanglement measures \cite{Cornfeld_2018,Feldman:2019upn,Chen_2021,chen2022negativityboson,parez22,multicharged,Wellnitz:2022cuf,operator_entanglement,fidelity}.
	Here, we continue to work in this direction by studying a system of one-dimensional massive Dirac fermions and extracting their symmetry resolved entropy at finite temperature and size. In the context of
	non-complementary subsystems and mixed-states, we also investigate the {\it charge imbalance resolved negativity} at finite temperature and size, generalising the results of \cite{Murciano_2021} by taking into account the mass correction. 
	
	The paper is organised as follows. In Section \ref{sec:defs} we provide all the definitions concerning the measures of symmetry resolved entanglement. We also introduce the concept of negativity for fermionic
	systems and we discuss the charge imbalance resolved negativity. This setup allows us to calculate the symmetry resolved entanglement entropy of massless Dirac fermions for a subsystem with multiple intervals at finite size and temperature in Section \ref{sec:masslessE}. By adding a massive perturbative term to the Dirac action, we  generalise the previous study in Section \ref{Section:massive_calculations}. The approach we adopt is suitable also to study the charge imbalance negativity, extending the results found in \cite{Murciano_2021} to the massive case in Section \ref{sec:fneg}. Throughout the manuscript, we check the agreement of the analytic expressions with the numerical lattice computations, both for the charged moments and their Fourier transforms. We draw our conclusions in Section \ref{sec:conclusion}. Five appendices provide the details of our computations, especially for the analytic continuations of the R\'enyi entropy.
	
	\section{Entanglement resolution in pure and mixed states}\label{sec:defs}
	In this section, we provide the general definitions of the main quantities of interest, namely the symmetry
	resolved entanglement entropies and the charge imbalance negativity: they quantify the entanglement in each charge sector when the system is invariant under a global $U(1)$ symmetry.
	A charge sector is defined as a $U(1)$ invariant subspace of the Hilbert space \cite{Goldstein_2018}.
	\subsection{Symmetry resolved entanglement}\label{sec:review_bosonization}
	Le us briefly review the study of critical systems with an additional internal global $U(1)$ symmetry. The generator of the symmetry, which we call $Q$, commutes with the Hamiltonian and thus the density matrix, that for a system at nonzero temperature $1/\beta$ reads $\rho=e^{-\beta H}/Z$, with $Z=\mathrm{Tr} e^{-\beta H}$.
	If the generator $Q$ can be written as the sum of the local charges in the two subsystems $A \cup B$, $Q=Q_A+Q_B$, the charge restricted to $A$ commutes with the reduced density matrix
	\begin{equation}
		[Q_A,\rho_A]=0,
	\end{equation}
	and the RDM takes a block form in the basis of the eigenstates of $Q_A$, with eigenvalues $q$:
	\begin{equation}
		\rho_A=\begin{pmatrix}
			p(q_1)\rho_A(q_1)&0&0&\ldots\\
			0&p(q_2)\rho_A(q_2)&0&\ldots\\
			0&0&p(q_3)\rho_A(q_3)&\ldots\\
			\vdots&\vdots&\vdots&\ldots\\
		\end{pmatrix}.
	\end{equation}
	The factors $p(q)$ have been introduced in order to normalise each block, so that \begin{align}
		\Tr(\rho_A(q))=1, \qquad
		\sum_q p(q)=1.
	\end{align}
	They describe the probability of finding $q$ as an outcome of the measurement of $Q_A$.
	The density matrix of each block can be found projecting $\rho_A$ in the corresponding eigenspace and normalising the result
	\begin{equation}
		\rho_A (q)=\f{\Pi_q \rho_A \Pi_q}{\Tr(\Pi_q \rho_A)},
	\end{equation}
	where $\Pi_q$ is the projector in the $q$-charge sector.
	Since each $\rho_A(q)$ is a density matrix (meaning it is positive semidefinite and normalised), the corresponding entanglement entropy, which we call symmetry resolved entanglement entropy, is given by
	\begin{equation}S_A(q)=-\Tr(\rho_A(q)\log(\rho_A(q))),\end{equation}
	and is related to the total entanglement entropy as
	\begin{align}\label{eq:sum}
		S_A=\sum_q [p(q) S_A(q)-p(q)\log(p(q))].
	\end{align}
	
	From this expression, we can decompose the total entanglement as a sum of two quantities. The first one is the weighted sum of the entanglement in each symmetry sector, which has been called \emph{configurational entanglement entropy} $S^c$ \cite{Lukin}; the second one, instead, arises from the fact that the system can be found in each symmetry sector with a certain probability $p(q)$. This uncertainty, which can be viewed as ``missing" information, is called \emph{number entanglement}, $S^{num}$.\\ One can also define the symmetry resolved R\'enyi entropy as
	\[S_n(q)=\f{1}{1-n}\Tra{\rho_A(q)^n}.\numberthis\]
	To compute these quantities using CFT, we will follow the approach described in \cite{Goldstein_2018}. Exploitng the property that the eigenvalues of $Q_A$ only take integer discrete values (for the $U(1)$ symmetry), we can  conveniently express the projector with a Fourier transform 
	\begin{align}
		\Pi_q=\int_{-\pi}^\pi e^{iQ_A\alpha} e^{-i q \alpha}\f{\d \alpha}{2\pi}.
	\end{align}
	By introducing the charged moments $Z_n(\alpha)=\mathrm{Tr}(\rho_A^n e^{i\alpha Q_A})$, the projected density matrix in a charge sector is a Fourier transform
	\begin{align}\label{eq:ft}
		\Op{Z}_n(q)\equiv \mathrm{Tr}(\Pi_q \rho_A^n)=\int_{-\pi}^\pi e^{- i \alpha q}Z_n(\alpha)\f{\d \alpha}{2\pi}, \quad \Op{Z}_1(q)=p(q),
	\end{align}
	and the symmetry resolved R\'enyi entropies read
	\begin{align}
		S_n(q)=\f{1}{1-n}\log \f{\Op{Z}_n(q)}{\Op{Z}^n(q)}, \quad S_1(q)=\lim_{n\to 1}S_n(q).\label{renyi entropy symmetry-resolved(Sela-G)}
	\end{align}
	Luckily, the charged moments can be computed via the replica trick; the factor $e^{i\alpha Q_A}$ is translated in a phase difference condition in the path integral formulation \cite{Goldstein_2018}. This means that, when sewing all the Riemann sheets together, we get a phase $e^{ i \alpha}$ whenever we complete a cycle through the $n$ copies.

	\subsection{Fermionic negativity and charge imbalance}\label{sec:details_neg}
	As mentioned in the introduction, the limit of applicability for entanglement entropy lies in the requirement that the system is in a pure state, thus excluding the case of  a finite temperature system or a multipartite geometry.
	In this subsection, we review the definition of fermionic negativity presented in \cite{Shapourian2_2017}, which covers such cases.
	Consider a one-dimensional subsystem $A$, described by a RDM $\rho_A$, partitioned in two parts $A_1$ and $A_2$.
	The fermionic negativity can be defined by looking at the action of a  \textit{partial time-reversal} transformation on the density matrix. This operation, in bosonic systems, is equivalent to a partial transposition, while it differs for a phase in the case of fermions.
	This property can be conveniently shown by looking at the time-reversal operation on fermionic coherent states \cite{Shapourian_2017}, which is:\begin{align}
		\left(\ketbra{\bar{\xi}}{\xi}\right)^R=\ketbra{i\xi}{i\bar{\xi}},
	\end{align}
	where $\xi,\bar{\xi}$ are Grassman variables and $\ket{\xi}=\mathrm{e}^{-\xi c^{\dagger}}\ket{0}$,$\bra{\bar{\xi}}=\bra{0}\mathrm{e}^{- c\bar{\xi}}$ are the related fermionic coherent states while $c, c^{\dagger}$ are the fermionic operators satisfying $\{c^{\dagger}_i,c_j\}=\delta_{ij}$.
	The factor $i$ is necessary for the anticommuting nature of the Grassman variables, and it is required to leave the trace invariant under this operation. The superscript $R$ is used to distinguish the fermionic from the standard partial transpose operation defined in Eq. \eqref{eq:bosonic}.
	This transformation rule  can be rewritten in the occupation number basis as 
	\begin{multline}
		\left(	\ketbra{\{n_i\}_{i\in A_1},\{n_j\}_{j\in A_2}}{\{\bar{n_i}\}_{{i}\in A_1},\{\bar{n}_j\}_{j\in A_2}}\right)^{R_1}=\\
		=
		\left(-1\right)^{\phi\left(\{n_i\},\{\bar{n}_i\}\right)}\left(	\ketbra{\{\bar{n_i}\}_{{i}\in A_1},\{\bar{n}_j\}_{j\in A_2}}{\{n_i\}_{i\in A_1},\{n_j\}_{j\in A_2}}\right)\label{eq13}
	\end{multline}
	where
	\begin{align}
		\phi\left(\{n_i\},\{\bar{n_i}\}\right)=\f{\tau_1+\bar{\tau_1}\mod 2}{2}+\left(\tau_1+\bar{\tau_1}\right)\left(\tau_2+\bar{\tau_2}\right)\quad \tau_x=\sum_{i\in A_x} n_i\quad\bar{\tau}_x=\sum_{i\in A_x} \bar{n}_i. 
	\end{align}
	It shows that the partial time-reversal acts as the composition of the partial transpose plus a unitary operator, i.e. a phase \cite{Shapourian_2017}:
	\begin{multline}
		\left(	\ketbra{\{n_i\}_{i\in A_1},\{n_j\}_{j\in A_2}}{\{\bar{n_i}\}_{{i}\in A_1},\{\bar{n}_j\}_{j\in A_2}}\right)^{R_1}=\\=U_A\left(	\ketbra{\{n_i\}_{i\in A_1},\{n_j\}_{j\in A_2}}{\{\bar{n_i}\}_{{i}\in A_1},\{\bar{n}_j\}_{j\in A_2}}\right)^{T_1}U_A\dagg.\label{eq15}
	\end{multline}
	In this basis, the phase contribution is $0$ along the diagonal, since $\phi\left(\{n_i\},\{{{n}_i}\}\right)=0$.\\
	Under the partial time-reversal $R_1$, the density matrix is no longer hermitian, in general. To restore this property, we can consider the matrix $\abs{\rho^{R_1}}=\sqrt{{\rho^{R_1}}\dagg\rho^{R_1}}$ which has only real eigenvalues, and the fermionic negativity is defined by
	\begin{align}
		\Nop=\f{\Tr \abs{\rho^{R_1}}-1}{2}.
	\end{align}	
	Analogously to R\'enyi entropy, we can define the \emph{fermionic Rényi negativity} as
	\begin{align}
		R_n=\begin{cases}
			\log \Tr\left(\rho^{R_1}{\rho^{R_1}}\dagg\right)^\f{n_e}{2} \qquad &n=n_e \quad\text{even},\\
			\log \Tr \left[ \left(\rho^{R_1}{\rho^{R_1}}\dagg\right)^\f{n_o-1}{2}\rho^{R_1}\right] \qquad &n=n_o \quad\text{odd}.\\
		\end{cases}\label{Renyi_negativity}
	\end{align}
	From Eq. \eqref{Renyi_negativity}, we can recover the fermionic negativity by the analytical continuation  of the \emph{even} Rényi negativities, evaluated in $n_e=1$ as
	\begin{align}
		\mathcal{\Nop}=\unm\lim_{n_e\ra 1}\left(R_{n_e}-1\right).
	\end{align}
	
	To further investigate the internal symmetry structure of this quantity in the presence of a conserved charge, we can follow Refs. \cite{Cornfeld_2018, Bonsignori_2019}: let us call $Q_A$ the conserved charged restricted to $A$, which in turn can be expressed as the sum $Q_1$ and $Q_2$ of charges localised in the two partitions of the system $A_1$ and $A_2$, respectively. The density matrix commutes with the total charge and we act on the commutation relation with a partial time-reversal operation, as in Eq. \eqref{eq15}: 
	\begin{align}
		&	0=[\rho_A,Q_A ]=[\rho_A,Q_1 ]+[\rho_A,Q_2 ]
		\implies\nonumber\\
		&	 0=[\rho_A,Q_2 ]+U_A\left([\rho_A,Q_1 ]\right)^T U_A\dagg =[\rho_A,Q_2 ]-[U_A\dagg\rho_A^{T}U_A,U_A\dagg Q_1^TU_A ]=[\rho_A^{R_1},Q_2-Q_1^{R}]
	\end{align}
	In the occupation number basis, which diagonalises the charge operator, we can see explicitly $Q_1^R=Q_1$, since $Q$ is diagonal in this basis, so the conserved charge of the partial time-reversed operator, $Q_{imb}=Q_2-Q_1$, has the form of a \emph{charge imbalance} between $A_1$ and $A_2$.\\
	Exploiting the presence of a conserved charge, we can decompose the total negativity of the system as done for the entropy:
	\begin{align}
		\Nop=\sum_q p(q) \Nop(q),
	\end{align}
	where the normalising factor $p(q)$ now is defined as \begin{align}
		p(q)=\Tr(\Pi_q \rho_A^{R_1}),
	\end{align}
	and $\Pi_q$ is the projector in the symmetry $q$ imbalance sector. 
	Let us remark that if our initial density matrix $\rho_A$ is in a pure state, there exist some degenerate values of $q$ for which $p(q)=0$; this is due to the fact that the value of $Q_A=Q_1+Q_2$ is fixed in this case, thus allowing only certain values of the imbalance $Q_{imb}=Q_2-Q_1$ to be different from zero. This is not the case for mixed states.\\
	Finally, we can define the charged moments of the partial time-reversed density matrix 
\begin{align}
N_n(\alpha)=
\begin{cases}
		\mathrm{Tr}\left( \left(\rho_A^{R_1}(\rho_A^{R_1})^{\dagger}\right)^n e^{i\alpha Q_{imb}}\right), \qquad \,\,\, n=2m	\\
		\mathrm{Tr}\left(\left(\rho_A(\rho_A)^{\dagger}\right)^{n-1} \rho_A e^{i\alpha Q_{imb}}\right), \quad n=2m+1
\end{cases}\label{momentsnegativity}
\end{align}
	and the charge imbalance resolved moments as their Fourier transform
	\begin{align}
		\Op{Z}_{R_1,n}(q)=\int_{-\pi}^\pi \d \alpha e^{-i \alpha q} N_{n}(\alpha). \label{eq:inverse Fourier}
	\end{align}
	From the moments, we can recover the normalised R\'enyi negativity 
	\begin{align}
		\Nop_n(q)=\f{\Op{Z}_{R_1,n}(q)}{p(q)^n} \qquad p(q)=\Nop_1(q)\label{momentstransform}.
	\end{align}
	The  analitycal continuation performed on the even numbers provides the charge imbalance negativity according 	to the limit 
	\begin{align}
		\mathcal{N}(q)=\lim_{n_e \to 1}\unm\left(\mathcal{N}_{n_e}(q)-1\right).
	\end{align}

	\section{Warmup: Symmetry resolution - massless case}\label{sec:masslessE}
	As a warmup, we consider the massless Dirac field theory, described by the following Lorentz invariant Lagrangian in a $1+1$ dimensional spacetime
	\begin{equation}
		\label{eq:lagrangianDirac}
		\mathcal{L}= \bar{\psi}\gamma^\mu \partial_\mu \psi ,
	\end{equation}  
	where $\bar{\psi}=\psi^\dagger \gamma^0$. The $\gamma^\mu$ matrices can be represented 
	in terms of the Pauli matrices as $\gamma^0=\sigma_1$ and $\gamma^1=\sigma_2$.
	This action has a global $U(1)$ symmetry, $\psi\mapsto e^{i \alpha}\psi$ 
	and $\bar{\psi}\mapsto e^{-i \alpha}\bar{\psi}$ which is related to the conservation of the charge $Q=\int {\rm d}x_1 \psi^\dagger\psi $.
	Here, we focus first on the computation of the charged moments through a diagonalisation in the space of replicas. Using the bosonisation technique, the problem is mapped to the calculation of a correlation function of vertex operators.
	The analytical expressions are compared with the lattice computations, finding a good agreement. Finally, we perform a Fourier transform of the obtained quantities. We think it is instructive to report all the steps for the derivation of the known results about the massless case since they will be the building blocks for the following sections about the massive Dirac fermions, whose computations are more cumbersome.
	
	\subsection{Replica diagonalisation}
	Let us consider a system of free one-dimensional Dirac fermions at finite temperature and size, whose corresponding field is defined on a torus spacetime, given the periodicity in both imaginary time $\tau$ and space $x$. 
	The partition function corresponding to $\Tra{\rho^n_A}$ has a single fermionic field defined on a Riemann surface made of $n$ different sheets, and can be mapped to an equivalent one in which one deals with a $n$-component field, which is instead defined on a single sheet:
	\begin{equation}
		\label{eq:nfieldsonC}
		\Psi=
		\begin{pmatrix}
			\psi_1
			\\
			\psi_2
			\\
			\vdots 
			\\
			\psi_n
		\end{pmatrix},
	\end{equation}
	where $\psi_j$ is the Dirac field of the $j$-th copy of the system.
	
	These copies interact through a \emph{composite twist field operator} $\mathcal{T}_{n,\alpha}$ which imposes the appropriate boundary conditions that, in the Riemann surface picture, connects the various sheets. Given a component of the field $\psi_j$, the insertion of a twist field at a point $u$ implies that a winding around such point
	$(z-u)\mapsto e^{2\pi i}(z-u)$ maps it to the next component, with an additional phase $e^{i\alpha/n}$.
	The same applies to the composite anti-twist field $\tilde{\mathcal{T}}_{n,\alpha}$, which takes a field from the copy $j$ to $j-1$
	adding a phase $e^{-i\alpha/n}$. This transformation can be encoded in the twist
	matrix:
	\begin{align}
		T_{\alpha}=\begin{pmatrix}
			0&-e^{i\f{\alpha}{n}}&0&\ldots&0	&0\\
			0&0&-e^{i\f{\alpha}{n}}&\ldots&0	&0\\
			\vdots&\vdots&\vdots&\vdots&\vdots&\vdots\\
			0&0&0&\ldots&0&-e^{i\f{\alpha}{n}}	\\
			e^{i\f{\alpha}{n}}&0&0&\ldots&0&0	
		\end{pmatrix}\qquad \Top_{n,\alpha}: \psi_i\ra \left[T_{\alpha}\right]_{ij}\psi_j\label{twistmatrix1}
		.\end{align}	
	The action of this matrix is diagonalised by the appropriate change of basis in the replica space:
	\begin{equation}\tilde{ \psi}_k\equiv\sum_{j=1}^n e^{-i\left(\f{2\pi k\cdot j}{n}-\pi j\right)}\psi_j,\quad
		T_\alpha \tilde{ \psi}_k=e^{i \f{\alpha+2\pi k }{n}}\tilde{ \psi}_k,
	\end{equation}
	for $k=-\f{n-1}{2},\dots \f{n-1}{2}$,
	so that the twist fields can be decomposed into $n$ different twists acting independently on each replica
	\[\Top_{n,\alpha}=\prod_k \Top_{k,n,\alpha},\qquad \tilde{\Top}_{n,\alpha}=\prod_k \tilde{\Top}_{k,n,\alpha} .\]
	Since the  $\tilde{\psi}_k$ fields are decoupled, the total partition function is just the product of $n$ independent partition functions, ${Z}_k$. In this formalism, the addition of a global flux $\alpha$ is straightforward by equally splitting it among all the $n$ replicas.\\
	There is an ambiguity in the choice of the phase as it can generically take the form $\f{2\pi k+\alpha}{n}+2\pi m$. Different choices of the winding number $m$ correspond to different configurations of the field, and in principle we should take a linear combination of the partition functions computed with all the possible choices of $m$.
	However, the leading order contribution to the partition function $Z_k$ corresponds to the choice $m=0$, as explained in Appendix \ref{fluxambiguity}.
	
	We now review the procedure outlined in \cite{Casini_2005} to compute the decoupled partition functions $Z_k$. The most generic subsystem $A$ is composed of $p$ intervals
	\[A=\bigcup_{i} [u_i,v_i],\numberthis \]
	the charged moments can then be expressed as a correlator
	\[Z_n(\alpha)=\ave{\prod_{i=1}^p{{\Top}}_{n,\alpha}(u_i){\tilde{\Top}}_{n,\alpha}(v_i)}=\prod_{k=-\f{n-1}{2}}^{\f{n-1}{2}}\ave{\prod_{i=1}^p{{\Top}}_{k,n,\alpha}(u_i){\tilde{\Top}}_{k,n,\alpha}(v_i)}\equiv \prod_k Z_k (\alpha), \label{partitionfunctiondecoupled}\numberthis \]
	where we indicate as $u_i,v_i$ the points corresponding to $\vec{u}_i=(u_i,0),\vec{v_i}=(v_i,0)$ in the Euclidean coordinates of the path integral.
	Our fields are multivalued: they gain a different phase around the branch points $u_i$ and $v_i$, respectively. In order to obtain single-valued fields, we can let the phase be absorbed by a gauge transformation \cite{Casini_2005}. 
	Calling $\psi^G$ the new, gauged, fields, they are related with the multivalued ones by the gauge transformation 
	\[\tilde{ \psi}_k(\vec{x})\ra \psi^G_k=\exp \left[ i\int_\vec{0}^\vec{x} \Aop_\mu(\vec{x}')\d x'^\mu \right]\tilde{ \psi}_k(\vec{x}). \numberthis \]
	Requiring a null phase condition around the branch points for $\psi^G_k$, Stokes theorem allows to find the appropriate gauge transformation:
	\begin{equation}
		\f{2\pi k+\alpha}{n}+\oint_{u_i} \Aop_\mu(\vec{x}')\d x'^\mu=0 \implies \epsilon^{\mu\nu} \partial_\nu\Aop_\mu(\vec{x}')=\f{2\pi  k+\alpha}{n}\delta^{(2)}(\vec{x}-\vec{u}_i).
		\label{Gaugefield definition}
	\end{equation}
	where $\epsilon_{\mu\nu}$ is the antisymmetric tensor in $2$ dimensions.
	Similarly, the points $\vec{v}_i$ give an opposite phase and putting all together we get
	\begin{equation}
		\epsilon^{\mu\nu} \partial_\nu\Aop_\mu(\vec{x})=\sum_i\f{2\pi k+\alpha}{n}\left(\delta^{(2)}(\vec{x}-\vec{u}_i)-\delta^{(2)}(\vec{x}-\vec{v}_i)\right).\label{Gauge potential}\end{equation}
	The Dirac Euclidean lagrangian in a gauge field is      
	\begin{equation}
		\Op{L}_k=\bar{\psi}^G_k\gamma^\mu(\partial_\mu-i\Aop_\mu)\psi^G_k+m{\bar{\psi}^G_k\psi^G_k}.\label{Dirac lagrangian}\end{equation}
	We can use the fermionic current $j^\mu=\bar{\psi}^G_k\gamma^\mu \psi^G_k$ to isolate the gauge field term in the action, rewriting the partition function as
	\[Z_k=\ave{e^{i \int j_\mu A^\mu\d^2 x}}_{CFT},\label{eq12}\numberthis \]
	where $\langle \cdot \rangle_{CFT}$ denotes that the expectation value is taken on the massless, ungauged theory, which is conformal invariant.
	We can apply the bosonisation technique, with the substitution \begin{align}
		j^\mu \rightarrow\f{1}{2\pi}\epsilon^{\mu\nu}\partial_\nu\phi_k ,\label{eq:bosonized_current}
	\end{align}
	and compute Eq. \eqref{eq12} in the dual theory of the Dirac fermion, which is that of a free scalar $\phi_k$, described by the action:
	\[S=\int\f{1}{8\pi}(\partial_\mu \phi_k)^2\d^2 x.\numberthis \]
	Because of Eq. \eqref{Gauge potential}, $Z_k(\alpha)$ is nothing but a correlation function of vertex operators:
	\begin{equation}\label{partitionfunctiondefinition}
		Z_k(\alpha)=\ave{\prod_je^{i\f{k+\alpha/2\pi}{n} \phi_k(u_j)}e^{-i\f{k+\alpha/2\pi}{n} \phi_k(v_j)}}.
	\end{equation}
	The correlation function of vertex operators on a torus is given by \cite{DiFrancesco:639405}:\begin{align}
		\ave{\prod_i \exp{i \alpha_i \phi(z_i)}}=\delta_{0,\sum_i {\alpha_i}}\: \abs{\frac{\theta_\nu\cbr{\sum_i \alpha_i z_i|\tau}}{\theta_\nu\cbr{0|\tau}} }\prod_{i<j} \abs{\frac{\partial_z\eva{\theta_1\cbr{z|\tau}}{z=0}}{\theta_1\cbr{z_i-z_j|\tau}}}^{-2 \alpha_i \alpha_j}.
		\label{eq:corrfunctiontorus}\end{align}
	Using Eqs. \eqref{eq:corrfunctiontorus} and \eqref{partitionfunctiondefinition}, we find:
	\begin{equation}
		Z_k(\alpha)=\Big|\f{\prod_{a<b}\theta_1(\frac{u_a-u_b}{L}|\tau)\theta_1(\frac{v_a-v_b}{L}|\tau)}{\prod_{a,b}\theta_1(\frac{u_a-v_b}{L}|\tau)}\left({\epsilon} \partial_z \theta_1(0|\tau)\right)^p\Big|^{2\f{(k+{\alpha/2\pi}{})^2}{n^2}}\left|\f{\theta_\nu(\sum_a\f{k+\alpha/2\pi}{n}(\frac{u_a-v_a}{L}))}{\theta_\nu(0|\tau)}\right|^2\label{kdependentpartitionfunction},
	\end{equation}
	where  we have added  an ultraviolet cutoff  $\epsilon\approx\frac{a}{L}$ necessary for the numerical comparison with a lattice theory with spacing $a\ra0$.
	The spacial length of the torus is set equal to $L$, and  $\tau=\frac{i\beta}{L}$  is the ratio of the two periods of the torus.
	$\theta_\nu$ are the Jacobi theta functions, with the pedix $\nu$ specifying one of the four possible boundary conditions along the torus, the \textit{spin sectors} $\nu=1,2,3,4$. Their meaning is reported in the following table using the standard notation of $NS$ and $R$ for Neveu-Schwarz or Ramond periodic condition, respectively:\begin{center}
		\begin{tabular}{ c c c }
			Spin sector & Periodicity in the $x$ direction & Periodicity in the $\tau$ direction \\ 
			$\nu=1$ & R & R\\ 
			$\nu=2$ & R & NS\\ 
			$\nu=3$ & NS & NS\\ 
			$\nu=4$ & NS & R
		\end{tabular}
	\end{center}
	\quad \\
	Before concluding this section, we notice that Eq. \eqref{kdependentpartitionfunction} can be put in another form by applying a modular transformation on the parameter $\tau\ra-\f{1}{\tau}$.
	Under such transformation, the Jacobi theta functions obey the identity  \cite{DiFrancesco:639405}:\begin{align}
		\theta_\nu(x|\tau)=-\frac{e^{-\frac{\pi  i  x^2}{\tau}}}{\sqrt{-i\tau}}\theta_{\nu'}\left(\frac{x}{\tau}|-\frac{1}{\tau}\right){},
	\end{align}
		where $\nu'=\nu$ for $\nu=1,3 $, while $\nu=2,4$ interchange.
		Applying this transformation to \eqref{kdependentpartitionfunction}, we obtain the expression:
	\begin{multline}
		Z_k(\alpha)=\Big |\f{\prod_{a<b}\theta_1(\f{u_a-u_b}{i\beta}|-\f{1}{\tau})\theta_1(\f{v_a-v_b}{i\beta}|-\f{1}{\tau})}{\prod_{a,b}\theta_1(\f{u_a-v_b}{i\beta}|-\f{1}{\tau})}\left(\f{\epsilon}{\tau} \partial_z \theta_1\left(0\left|-\f{1}{\tau}\right.\right)\right)^p\Big|^{2\f{(k+{\alpha/2\pi}{})^2}{n^2}}\\
		\times \left|\f{\theta_{\nu'}(\sum_a\f{(k+\alpha/2\pi)(u_a-v_a)}{i\beta n}\left|-\f{1}{\tau}\right.)}{\theta_{\nu'}(0|-\f{1}{\tau})}\right|^2\label{kdependentpartitionfunctionmodulartransformed},
	\end{multline}

	\subsection{Charged moments on the torus}
	We can start from Eq. \eqref{partitionfunctiondecoupled} to compute the charged moments. To this end, we divide it in two parts, where the first factor, $Z_k^{0}(\alpha)$, does not depend on the spin sector $\nu$, so we dub it the \textit{spin independent part}. The second factor will be named instead the \textit{spin dependent part} $Z_k^\nu(\alpha)$, such that \begin{align}
		\log(Z_n)=\log(Z^\nu_n)+\log(Z^0_n )\label{eq:total_partition_function}
	\end{align}
	Explicit analytic expressions are provided in Eqs. \eqref{eq:chm}, \eqref{eq:nu2spindeppart}, \eqref{eq:nu3spindeppart}.
	\subsubsection{Spin independent part}\label{Spin independent Part}
	For convenience, we take the logarithm of both terms in \ref{partitionfunctiondecoupled}, to transform the product into a sum:
	\begin{align}
		\log Z_n(\alpha)=&\sum_{k=-\f{n-1}{2}}^{\f{n-1}{2}}2\f{(k+{\alpha/2\pi}{})^2}{n^2}\log \Big|\f{\prod_{a<b}\theta_1(\f{u_a-u_b}{i\beta}|-\f{1}{\tau})\theta_1(\f{v_a-v_b}{i\beta}|-\f{1}{\tau})}{\prod_{a,b}\theta_1(\f{u_a-v_b}{i\beta}|-\f{1}{\tau})}\nonumber\\&\times \left(\f{\epsilon}{\tau} \partial_z \theta_1\left(0\left|-\f{1}{\tau}\right.\right)\right)^p\Big|+\sum_{k=-\f{n-1}{2}}^{\f{n-1}{2}}\log\Big|\f{\theta_{\nu'}(\sum_a\f{(k+\alpha/2\pi)(u_a-v_a)}{i\beta n}\left|-\f{1}{\tau}\right.)}{\theta_{\nu'}(0|-\f{1}{\tau})}\Big|^2.\label{partitionfunctionnu for integers}\end{align}
	We focus on the physical sectors $\nu=2,3$ which are antiperiodic in the imaginary time with period $\beta$.
	The sum of the spin independent part can be easily calculated since
	\[ \f{2}{n^2}\sum_{k=-\f{n-1}{2}}^{\f{n-1}{2}}{(k+{\alpha/2\pi}{})^2}=\f{n^2-1}{6n}+\f{2(\alpha/2\pi)^2}{n} \numberthis, \]
	and we get
	\begin{multline}
		\label{eq:chm}
		\log Z^{0}_n(\alpha)=\\ =\left(\f{n^2-1}{6n}+\f{\alpha^2}{2\pi^2 n }\right)
		\log\left|\f{\prod_{a<b}\theta_1(\f{u_a-u_b}{i\beta}|-\f{1}{\tau})\theta_1(\f{v_a-v_b}{i\beta}|-\f{1}{\tau})}{\prod_{a,b}\theta_1(\f{u_a-v_b}{i\beta}|-\f{1}{\tau})}\left(\f{\epsilon}{\tau} \partial_z \theta_1\left(0\left|-\f{1}{\tau}\right.\right)\right)^p\right|.
	\end{multline}
	This expression, in the limit ${L,\beta \ra\infty}$, agrees with the well known CFT results  on a flat space-time \cite{Goldstein_2018} .

	\subsubsection{Spin dependent part}\label{Spin dependent Part}
	The spin dependent part of the partition function is the most interesting, since it shows deviations from standard CFT on a plane. As we will find out, these terms, which come purely from the boundary conditions, are responsible for small violations of the equipartition of entanglement. Physically, the influence of the boundary on entanglement shows the non-local nature of the latter.
	The goal is the computation of
	\begin{equation}
		\log Z_n^{\nu}(\alpha)=\sum_{k=-\f{n-1}{2}}^{\f{n-1}{2}}2\log\left|\f{\theta_{\nu'}(\f{(k+\alpha/(2\pi))r}{i\beta n}\left|-\f{1}{\tau}\right.)}{\theta_{\nu'}(0|-\f{1}{\tau})}\right|,
		\label{nu2sectorspindependentpartitionfunction}
	\end{equation}
	where $r$, from now on, will be defined as the total length of the subsystem: 
	\[r=\sum_a {u_a-v_a}. \numberthis \]
	Let us focus on the{ $\nu=2$} spin sector. Substituting the infinite product expansion of theta functions \cite{DiFrancesco:639405}, Taylor-expanding the logarithm and using the results for the geometric series, we find 
	\[\log(Z_n^{\nu=2})=2\sum_{m=1}^\infty \f{1}{m}\f{1}{\sinh(\frac{\pi m L }{\beta})} {\left(n-\f{\sinh(\f{\pi m r}{\beta})}{\sinh(\f{\pi m r}{\beta n })}\cosh(\f{ m r\alpha}{\beta n} )\right)}. \label{eq:nu2spindeppart}\numberthis \]
	This series converges exponentially fast for all real values of $n$. \\
	A similar formula can be found	for the spin sector $\nu=3$
	\[\log Z^{\nu=3}_n(\alpha)=2\sum_{m=1}^\infty \f{(-)^m}{m\sinh(\f{\pi m L}{\beta})} \left(n-\f{\sinh(\f{\pi m r}{\beta })}{\sinh(\f{\pi m r}{\beta n})}\cosh(\f{ m r\alpha}{\beta n} )\right).\label{eq:nu3spindeppart}\numberthis \]
	\subsubsection{Comparison with the lattice theory}
\begin{figure}[h]		
		\includegraphics{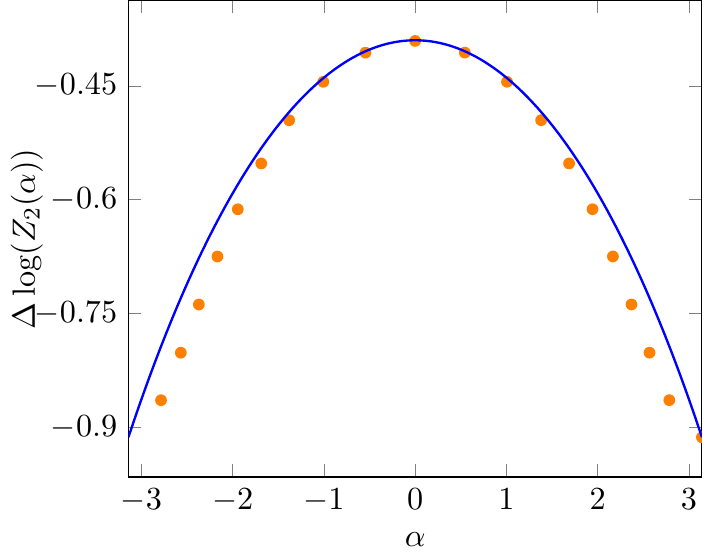}
		\includegraphics{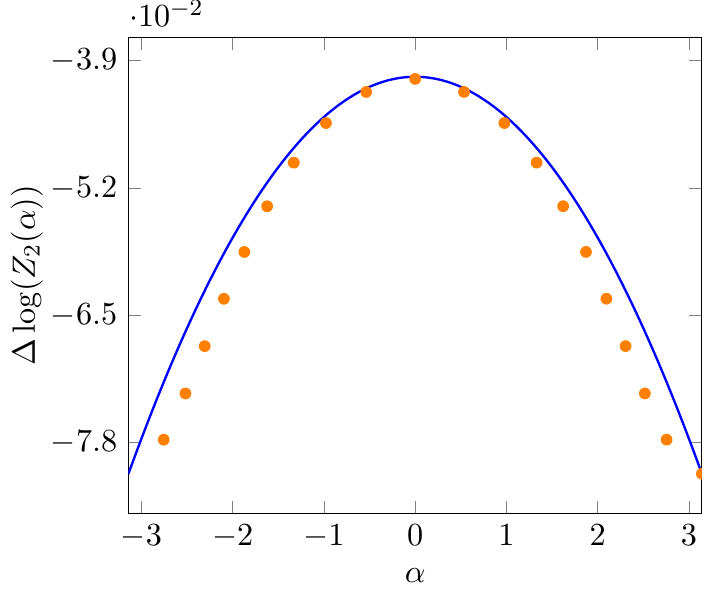}			
		\caption{Difference between the logarithm of the charged moments $\Delta\log Z_n(\alpha)\equiv \eva{\log Z_n(\alpha)}{r=\ell_1}-\eva{\log Z_n(\alpha)}{r=\ell_2}$, evaluated for a single interval of length, respectively, $\ell_2=0.3 L$ and $\ell_1=0.6 L$, in the spin sectors $\nu=2$ (left) and $\nu=3$ (right), with the temperature set to $\beta=3L$ and $N=300$ lattice sites. The blue continuous line uses the field theory prediction in Eq. \eqref{eq:total_partition_function} while the symbols correspond to the exact results obtained through the techniques reported in Appendix \ref{app:num}.
		}\label{diffnu2}
	\end{figure}

\begin{figure}
		\includegraphics{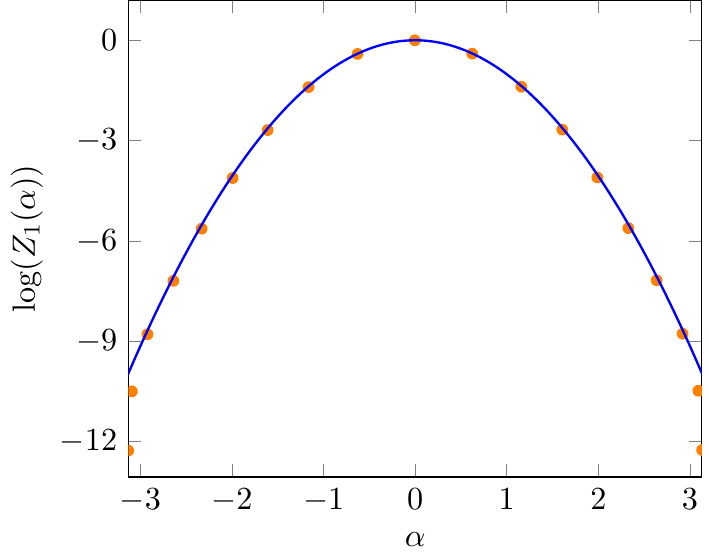}
		\includegraphics{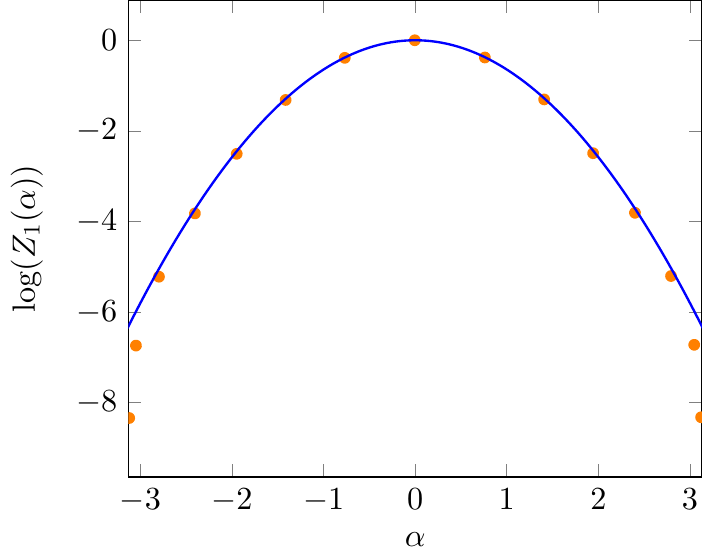}					\caption{Logarithm of the charged moment $\log(Z_1(\alpha))$ evaluated for a subsystem made of two intervals of length $0.1 L,\, 0.3 L$, separated by a distance $\Delta x =0.2 L$, at $\beta=0.1L$. The blue solid line corresponds to Eq. \eqref{partitionfunctionnu for integers} while the lattice cutoff $\epsilon$ is given by Eq. \eqref{eqepsilon}. We have set the number of lattice sites $N=300$.}
		\label{plot:alphadependencechargedmoments}
	\end{figure}

	In this section we benchmark the results for the charged moments $\log Z_n\equiv\log Z_n^{0}(\alpha)+\log Z_n^{\nu}(\alpha)$, $\nu=2,3$. For the lattice computations, we use the techniques reported in the Appendix \ref{app:num}, setting the number of lattice sites in the numerics to $N=300$. We also choose the lattice spacing $a=1/N$ in such a way that $L=Na=1$.\\
	As can be seen from Eq. \eqref{eq:chm}, there is an unknown lattice parameter, $\epsilon$, necessary to compare the CFT results with the numerics. To avoid this problem, a possibility is to consider quantities such as the difference of the logarithm of two charged moments, each computed for the same number of intervals of different length, so that the dependence on $\epsilon$ cancels out. This has been done in Figure \ref{diffnu2}, showing a good agreement between the lattice computations and the field theory prediction in Eqs. \eqref{eq:chm} and \eqref{nu2sectorspindependentpartitionfunction}.
	\\	It is also possible to fix $\epsilon$ using its analytical expression found in \cite{Bonsignori_2019} by exploiting the generalised Fisher-Hartwig conjecture.
	In short, this technique allows to obtain an explicit formula for the charged moments computed on a lattice theory at $T=0$ and $N\ra\infty$. The comparison with the analytical formula derived from field theory provides an expression for the lattice constant $\epsilon$, which we assume to weakly depend on $\tau=i\frac{\beta}{L}$.
	Numerical evidence seems to validate this assumption, as shown in Fig. \ref{plot:alphadependencechargedmoments}, where the lattice data and the field theory prediction perfectly overlap.
	
	We report explicitly the expression we used for the lattice constant $\epsilon$  \cite{Bonsignori_2019}, which is
	\begin{equation}
		\begin{split}
			\epsilon(n,\alpha)=&\f{1}{2 N} \exp\left(\f{6n}{n^2-1+12 (\alpha/2\pi)^2}\Upsilon(n,\alpha)\right),\\
			\Upsilon(n,\alpha)=& ni\int_{-\infty}^\infty \d w \left[\tanh( \pi w)-\tanh( \pi n w+{\unm i \alpha})\right] \log\left(\f{\Gamma\left(\unm+i w\right)}{\Gamma\left(\unm-i w\right)}\right).
		\end{split}\label{eqepsilon}
	\end{equation}

	\subsection{Symmetry resolved moments via Fourier transform}
\begin{figure}[h]			
	\includegraphics{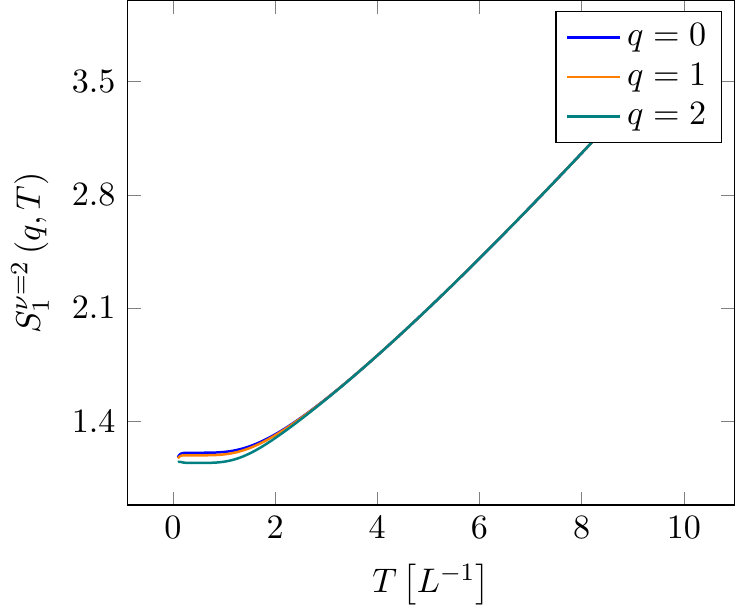}
		\includegraphics{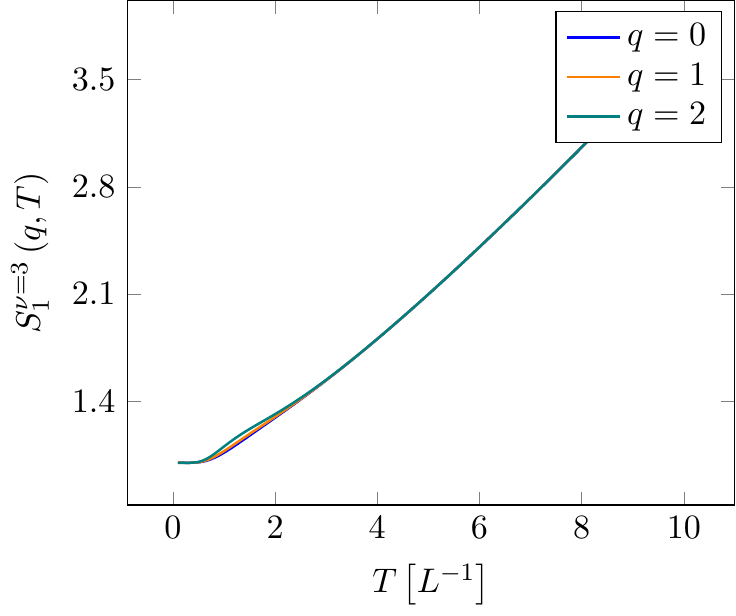}				\caption{Temperature dependence, in unit of the inverse spacial dimension of the system, for the entanglement entropy of a subsystem composed of an interval of length $0.4 L$ in the $\nu=2$ (left) and $\nu=3$ (right) spin sectors, obtained from Eq. \eqref{symmetry-resolved entropy saddle approximation}, with the constant $\epsilon$ evaluated for $N=300$ in Eq. \eqref{eqepsilon}. The asymptotic linear growth for higher temperatures reproduces the well known thermodynamic entropy of free fermions (quantum correlations are subleading at higher temperatures). }
		\label{Tdependence}
	\end{figure}

\begin{figure}
		\includegraphics{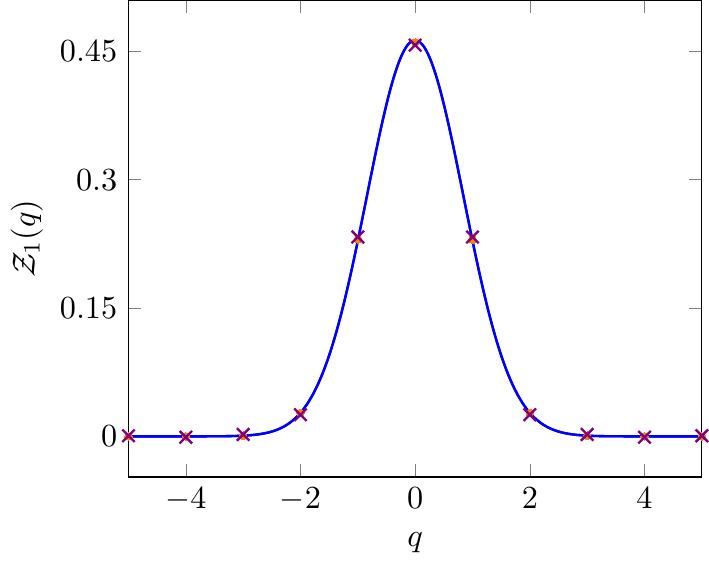}
		\includegraphics{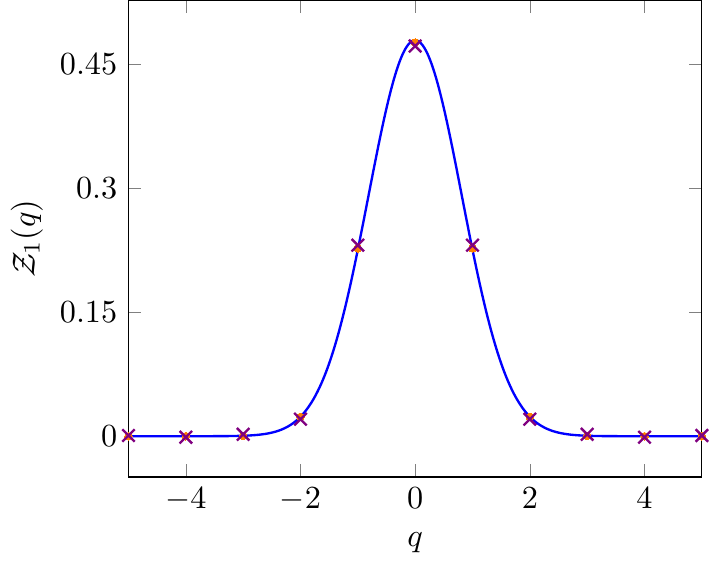}			\includegraphics{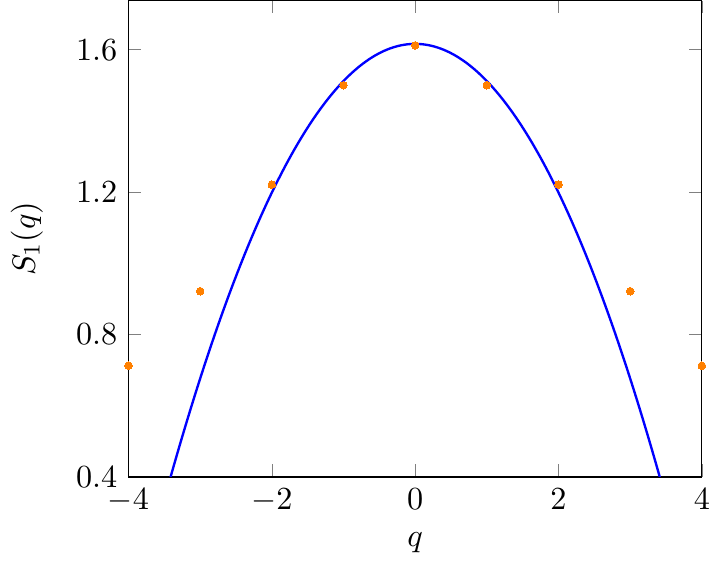}
		\includegraphics{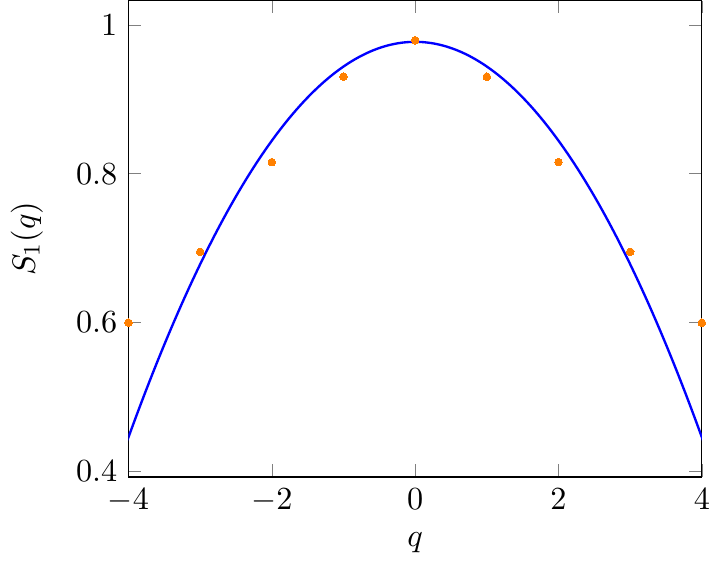}		
		\caption{Top panels: Symmetry resolved moments $\Op{Z}_n(q)$ for $n=1$, computed with the saddle point approximation in Eq. \eqref{saddle} (blue continuous line), with a numerical Fourier transform of the field theory charged moments in Eq. \eqref{partitionfunctionnu for integers} (cross mark) and with the lattice model (orange dots), for $N=300$. The values of the parameters are $\beta=10L$ and $r=0.3 L$, where $r$ is the size of the interval defining the subsystem. The plot on the left is obtained in the spin sector $\nu=2$, the other one corresponds to $\nu=3$. Bottom panels: Symmetry resolved entanglement entropy in the spin sector $\nu=2$ (left) and $\nu=3$ (right) for a subsystem made of a single interval of length $0.3L$, at $\beta=10 L$.
			The blue continuous line represents the result obtained from the field theory with the saddle point approximation in Eq. \eqref{symmetry-resolved entropy saddle approximation} and the orange dots show, instead, the numerical result for a finite lattice model ($N=300$).}
		\label{plot:masslessSREE}
	\end{figure}

	Applying the inverse Fourier transform \eqref{eq:inverse Fourier} we obtain the symmetry resolved R\'enyi entropies. In order to explicitly carry out this integral, we consider the physically relevant limit $\epsilon\sim\frac{a}{L}=\frac{1}{N}\ll 1$. The function $\log Z_n(\alpha)$ is well described by a quadratic function in $\alpha$ in this limit, since the spin independent part is dominant, thanks to the term $\propto \alpha^2 \log(\epsilon)$ in Eq. \eqref{eq:chm}: \begin{align}
		\abs{\log(Z^0_n)}\sim \left(\frac{n^2-1}{6n}+\frac{\alpha^2}{2 n \pi^2}\right)\log(N)\gg \abs{\log(Z^\nu_n)}=O(1).
	\end{align}
	In this limit, we can approximate the charged moments at the lowest nontrivial order in $\alpha$ as
	\[Z_n(\alpha)\approx A_n e^{-\f{b_n\alpha^2}{2}}, \label{eq:saddlepoint expansion}\numberthis \]
	where the explicit expression for the parameters $A_n, b_n$ are reported in Appendix \ref{app:saddlepoint} to lighten the notations.
	The Fourier transform can be explicitly carried out by extending the integration domain of Eq. \eqref{eq:ft} to the whole real axis, instead of just $[-\pi,\pi]$. This is a good approximation provided again $\f{b_n}{8}\gg 1$, which holds true in the thermodynamic limit $\epsilon\sim\frac{a}{L}\ra 0$.
	Within this assumption, we get the symmetry resolved moments
	\[\Op{ Z}_n(q)={A_n}\sqrt{\f{1}{2\pi b^{\nu}_n}} e^{-\f{( q)^2}{2 b^{\nu}_n}}, \label{saddle}\numberthis \]
	and the symmetry resolved entanglement entropy
	\begin{equation}
		S^{\nu}_n(q)=\f{1}{1-n}\left[\log \f{A_n}{A^n_1}+\unm\log \f{ (b^{\nu}_1)^n}{b^{\nu}_n} - \f{1-n}{2}\log(2\pi) -	\f{( q)^2}{2}\left(\f{n}{b^{\nu}_1}-\f{1}{b^{\nu}_n}\right)\right].\label{symmetry-resolved entropy saddle approximation}\end{equation}	
	It is interesting to study whether the equipartition of entanglement is broken. In particular, looking at Eq. \eqref{symmetry-resolved entropy saddle approximation}, the charge dependent contribution cancels if $b_n\propto\f{1}{n}$. Since this is always true for the spin independent part of the partition function (ignoring the $n$ and $\alpha$ dependence of the lattice constant $\epsilon$), any breaking of equipartition predicted by pure CFT is due to the spin dependent part of the charged moments, thus to the boundary conditions. Moreover, if we consider the explicit expression for $b_n^{\nu=3}$ in Eq. \eqref{divergnu3},  we observe that the spin dependent part is suppressed for $\frac{\beta}{L}\ll 1$, meaning the equipartition is preserved at high temperatures.
	\\ In Figure \ref{Tdependence}, we show the temperature dependence of the symmetry resolved entanglement entropy for one interval and different charges $q$, in both relevant spin sectors. We can clearly see that the equipartition is recovered in the high temperature regime, while this is spoiled by the boundary terms at low temperature, in particular in the spin sector $\nu=2$.
	In Figure \ref{plot:masslessSREE} (top panels), we plot the symmetry resolved moments obtained with the saddle point approximation in Eq. \eqref{saddle}, testing the agreement with the ones obtained from a numerical transform of the field theory formula in Eq. \eqref{partitionfunctionnu for integers} and with the lattice theory. Finally, in Figure \ref{plot:masslessSREE} (bottom panels) we do the same for the symmetry resolved entropy in Eq. \eqref{symmetry-resolved entropy saddle approximation}. We see that the agreement is good around $q=0$ and it worsens for higher values of $q$ since, while the lattice entropy seems to saturate, the saddle point predicts a parabolic behavior for all $q$, which only mimics the lattice behavior in a neighborhood of $q=0$.
	
	\section{Symmetry resolution - massive case}\label{Section:massive_calculations}
	
	In this section we derive the leading correction to the symmetry resolved entanglement entropy on the torus due to a mass term in the Dirac Lagrangian, i.e. 
	\begin{equation}
		\Op{L}=\bar{\psi}\gamma^\mu(\partial_\mu)\psi+m{\bar{\psi}\psi}.\label{Dirac lagrangian2}\end{equation}
	The procedure to compute the partition function is the same as the one described in Section \ref{sec:review_bosonization}, with the only difference that the average \eqref{eq12} is now computed on a massive theory and not on a CFT.
	Given the complexity of the resulting expressions, we refer to Appendix \ref{analytical_sum_ssEE} for the technical details, while we report the steps for their derivation in the main text. We show that the presence of a small mass in the Dirac action does not spoil equipartition at leading order, in the thermodynamic limit, but gives a subleading, spin sector dependent correction, which is proportional to the mass.
	
	Here, we treat the mass perturbatively, in such a way that the extra term in the Lagrangian is considered as a perturbation of the massless theory, as follows
	\begin{multline}
		Z_k^m(\alpha)=\ave{e^{i\int \d^2x\bar{\psi}\slashed{A}\bar{\psi}}}_{\text{massive theory}}\equiv\f{\int [D\bar{\psi}_k][D\psi_k] e^{-\int \d^2 x\bar{\psi}\left(\slashed{\partial}-i\slashed{A}+m\right)\psi}}{\int [D\bar{\psi}_k][D\psi_k] e^{-\int \d^2 x\bar{\psi}\left(\slashed{\partial}+m\right)\psi}}= \\ \label{eq19}
		=\f{\ave{ e^{\int \d^2 x\bar{\psi}\left(i\slashed{A}-m\right)\psi}}_{CFT}}{\ave{e^{-\int \d^2 x\bar{\psi}m\psi}}_{CFT}}. 
	\end{multline}
	We use the superscript $m$ in order to distinguish the massive charged moments from the massless ones. Using the bosonisation technique, Eq. \eqref{Dirac lagrangian2} can be mapped into a sine-Gordon model, whose action reads
	\[S_{SN}=\int \left[\f{1}{8\pi}\partial_\mu\phi\partial^\mu\phi+\lambda\cos(\phi) \right]\d^2 x,\numberthis \]
	where the coupling $\lambda$ is proportional to the mass with a dimensional factor dependent on a cutoff scale (this constant will be fitted when comparing the field theory and the numerical results).
	\\Eq. \eqref{eq19} can be rewritten in terms of expectation values on a bosonic theory replacing the massive term with the cosine interaction, and  using Eq. \eqref{eq:bosonized_current} for the current:
	\[Z^m_k(\alpha)=\f{\ave{e^{\int \f{\d^2 x }{2\pi} \left(i{A}_\mu\epsilon^{\mu\nu}\partial_\nu \phi-\lambda \cos(\phi)\right)}}}{\ave{e^{-\int \d^2 x\lambda \cos(\phi)}}},\numberthis \]
	where for brevity we omit the subscript CFT on the average from now on, assuming all path integrals are performed on the massless action.
	The first nontrivial order for the mass term is quadratic in $\lambda$, due to the neutrality condition of vertex operators:
	\[\ave{e^{-\int \d^2 x\lambda \cos(\phi)}}= \ave{1+\f{\lambda^2}{4} \int e^{i(\phi(x)-\phi(y))}\d^2 x\d^2 y+O(\lambda^4)}.\numberthis \]
	We can then express the first order mass correction to the charged moments as:
	\begin{equation}
		\begin{split}
			Z^m_k(\alpha)=\f{\ave{\prod_j e^{i\f{k+\alpha/2\pi}{n}\phi(0,u_j)}e^{-i\f{k+\alpha/2\pi}{n}\phi(0,v_j)}\left[1+\f{\lambda^2}{4} \int e^{i(\phi(x)-\phi(y))}\d^2 x\d^2 y\right]}}{\ave{1+\f{\lambda^2}{4} \int e^{i(\phi(x)-\phi(y))}\d^2 x\d^2 y}}
			\\
			\approx Z_k(\alpha)+\f{\lambda^2}{4}\ave{\left[\prod_j e^{i\f{k+\alpha/2\pi}{n}\phi(0,u_j)}e^{-i\f{k+\alpha/2\pi}{n}\phi(0,v_j)}-1\right]\int e^{i(\phi(x)-\phi(y))}\d^2 x\d^2 y}. 
		\end{split}
	\end{equation}
	We use the following compact notation to label our main quantities of interest:
	\begin{align}
		A_{k,n}(\alpha,x,y) \equiv&\f{\ave{e^{i(\phi(x)-\phi(y))}\prod_j e^{i\f{k+\alpha/2\pi}{n}\phi(0,u_j)}e^{-i\f{k+\alpha/2\pi}{n}\phi(0,v_j)}}}{\ave{\prod_j e^{i\f{k+\alpha/2\pi}{n}\phi(0,u_j)}e^{-i\f{k+\alpha/2\pi}{n}\phi(0,v_j)}}},\qquad \\ A^{tot}_n(\alpha)=&\sum_{k=-\f{n-1}{2}}^{\f{n-1}{2}} A_{k,n}(\alpha,x,y)-A_{k=0,n}(0,x,y),\label{eq:rego}
	\end{align}
	allowing us to write the correction to the full partition function as:
	\begin{equation}
		\begin{split}\label{eq:toderive}
			Z^{m}_n(\alpha)\approx Z^{m=0}_n(\alpha)\left(1+m^2\int A^{tot}_n(\alpha) \d^2 x\d^2 y\right) +O(m^4),
		\end{split}
	\end{equation}
	where, we stress again,  $\lambda \propto m$  up to a renormalisation constant. \\ $A_{k,n}$ is a correlation function of vertex operators, so Eq. \eqref{eq:corrfunctiontorus} yields:
	\begin{equation}
		\begin{split}
			A_{k,n}-\eva{A_{k,n}}{k=\alpha=0}&=\left|\f{\theta_\nu\left(\f{(k+\alpha ) r}{n}+x-y\right)|\tau}{\theta_\nu\left(\f{(k+\alpha ) r}{n}|\tau\right)}\f{\epsilon\partial_z\theta_1\left(0|\tau\right)}{\theta_1\left(y-x|\tau\right)}\right|^2
			\\	&\times \prod_a\left|\f{\theta_1\left(v_a-x|\tau\right)\theta_1\left(u_a-y|\tau\right)}{\theta_1\left(v_a-y|\tau\right)\theta_1\left(u_a-x|\tau\right)}\right|^\f{2(k+\alpha )}{n}-\left|\f{\epsilon\partial_z\theta_1\left(0|\tau\right)}{\theta_1\left(y-x|\tau\right)}^2\f{\theta_\nu(x-y|\tau)}{\theta_\nu(0|\tau)}\right|^2.\label{kdependent massive correction to the partition function}
		\end{split}
	\end{equation}
	The second addendum ensures that there is no mass correction when the size of the subsystem goes to $0$, and its presence is crucial to regularise the integral. Indeed, we can check that the integral in Eq. \eqref{kdependent massive correction to the partition function} does not diverge, so that a regularisation for the mass is not required to obtain a meaningful result. Assuming to work at finite $L,\beta$, a divergent contribution can only come from poles of the integrand. Since $\theta_1(z)\propto z$ as $ z\ra 0 ,$  $x=y$ is a singular point. Expanding around $x=y$ at the lowest order, we have
	\[\prod_a \left|\f{\theta_1\left(v_a-x|\tau\right)\theta_1\left(u_a-y|\tau\right)}{\theta_1\left(v_a-y|\tau\right)\theta_1\left(u_a-x|\tau\right)}\right|^\f{2(k+\alpha )}{n}=1+O(|x-y|), \numberthis \]
	and 
	\[\f{\theta_\nu\left(\f{(k+\alpha ) r}{n}+x-y|\tau\right)}{\theta_\nu\left(\f{(k+\alpha ) r}{n}|\tau\right)}=1+O(|x-y|), \numberthis \]
	which means that the two singular addends in Eq. \eqref{kdependent massive correction to the partition function}  cancel out at the leading order, and the divergence can behave at most as $\sim \f{1}{|x-y|}$. Since $x,y$ are $2-$dimensional variables, when divided in real and imaginary part, a pole must diverge at least like $\sim\f{1}{|x-y|^2}$ to give an infinite contribution, which is not the case.
	Points where $x=u_a,v_a$ or $y=u_a,v_a$ (and vice-versa) behave as $\abs{x-u_a}^{-\f{2(k+\alpha/(2\pi))}{n}}$. In our case, $|k|\le \f{n-1}{2}$ and $|\alpha|/(2\pi)\le \unm$, so this exponent can be $2$ at most, where the value $2$ occurs only for the most extreme values of $\alpha=\pm\pi$. Except for these $\alpha=\pm\pi$, such poles behave like $1/|x-x_0|^s,  s< 2 $
	which means they do not give a divergent contribution. 
	The analytical resummation of $\sum_{k}A_{k,n}$, can be done using  identities for the Jacobi theta functions and we report the result of these computations in Appendix \ref{analytical_sum_ssEE}.

	\subsection{Numerical implementation}
\begin{figure}[h!]
		\includegraphics{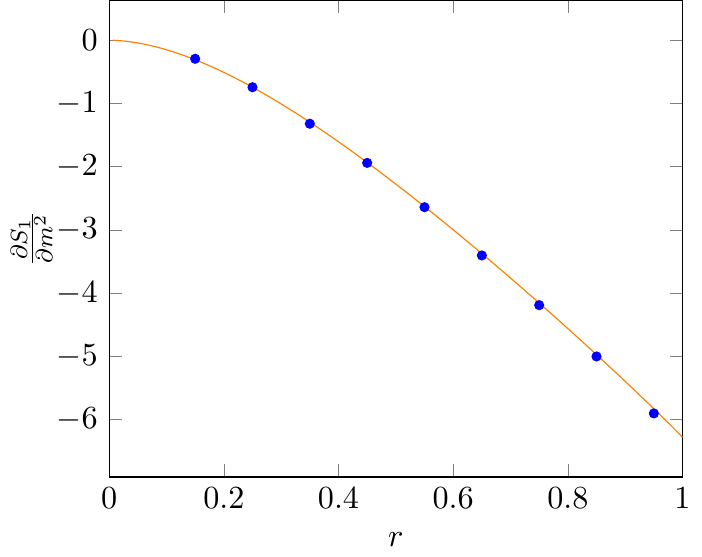}
		\includegraphics{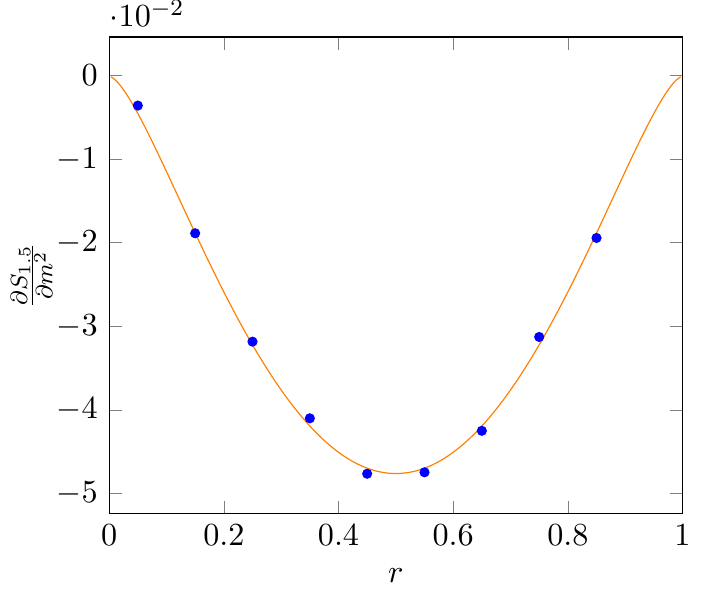}			\caption{Mass correction to the R\'enyi entropy as a function of the subsystems size, $r$ for $N=300$ sites. We compare the lattice results (continuous orange line) with Eq. \eqref{eq:entropy perturbation} (blue points) for a subsystem of one single interval, at $\frac{\beta}{L}=5$, in the spin sectors $\nu=2$ (left) and $\nu=3$ (right).
			We plot the R\'enyi entropy $S_n$ corresponding to $n=1$ for the $\nu=2$ spin sector, while we choose $n=1.5$ for the $\nu=3$ spin sector.	
			Notice that in this case we used the symbols for the field theory prediction, since the evaluation of the integrals defining this correction makes it hard to sample many points; for visual clarity, we used  instead a continuous line for the lattice data. The overlap between the field theory and the lattice theory is obtained by a fit of a multiplicative constant for the data, due to a renormalisation of the mass. }
		\label{plot:r-dependencemassivecorrectionRE}
	\end{figure}

	While the total entanglement entropy  at finite temperature and sistem size has already been studied in Ref. \cite{Herzog_2013} for the massless case, the massive correction was only found for integer values of $n\ge2$, so it is important to test the analytic continuation for $n=1$ and $\alpha=0$ against lattice computations.
	As already stressed, the mass is defined up to a renormalisation factor; to match the field theory result with the lattice model we fit a multiplicative constant for the renormalised mass and use it in the plots.
	We assume that the renormalised mass does depend on both the number of lattice sites and the ratio $\frac{\beta}{L}$, so a different fit is performed for every choice of these parameters.
	The R\'enyi entropies get a small mass contribution equal to
	\[\delta S^m_n=m^2\f{1}{1-n}\int A^{tot}_n \d^2 x \d^2 y,\label{eq:entropy perturbation}\numberthis \]
	and taking the limit $n\ra1$ of this expression (as explained in Appendix \ref{analytical_sum_ssEE}), we get the entanglement entropy mass correction, which has been plotted in Fig. \ref{plot:r-dependencemassivecorrectionRE}. 
	
	As noted in \cite{Herzog_2013}, since the ground state in the spin sector $\nu=2$ is degenerate, the limit $e^{-\beta H}|_{\beta\ra\infty}$ does not produce a pure state.
	The presence of the mass should remove this degeneracy, but our perturbative assumption requires $m\ll T,L^{-1}$, meaning that the degeneracy is not lifted and the limits $m\ra 0,\ T\ra 0$ do not commute. This is shown in Figure \ref{plot:r-dependencemassivecorrectionRE}, where, within the spin sector $\nu=2$, the property \begin{align}
		S_n(r)=S_n(L-r),
	\end{align} which holds for a pure state, is not respected.

	\begin{figure}
		\includegraphics{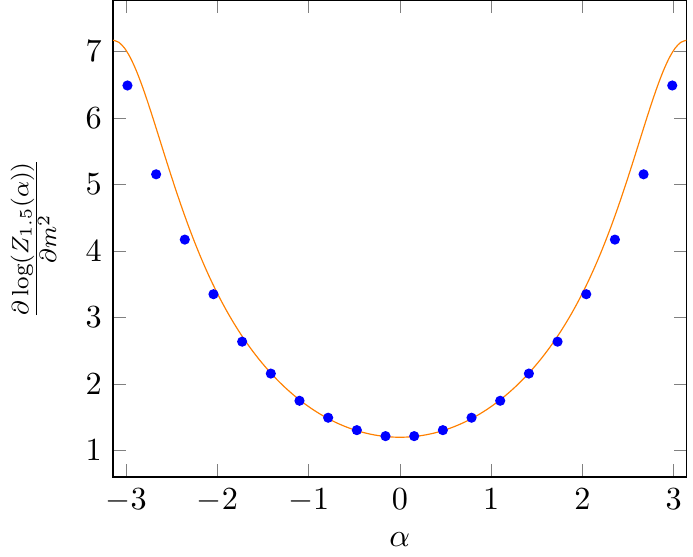}
		\includegraphics{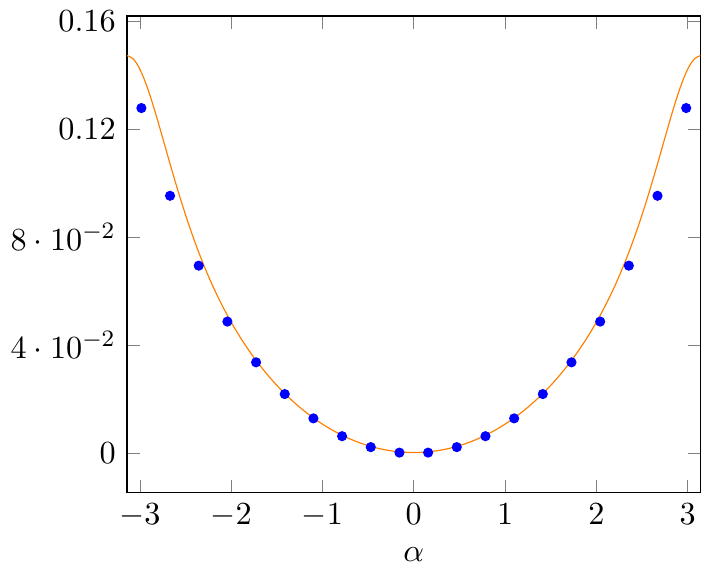}			
		\caption{Mass correction to the logarithm of the charged moment $\log(Z^m_n(\alpha))$ computed at $\beta=5$ for a subsystem of size $r=0.5 L$ in a system of $N=300$ sites. This quantity is evaluated for $n=1.5$, in the spin sector $\nu=2$ (left), and for $n=1$ in the spin sector $\nu=3$ (right).
			The plots have been overlapped thanks to a multiplicative constant  obtained by a subsystem-size fit like the one shown in Figure \ref{plot:r-dependencemassivecorrectionRE}. In particular, the blue points correspond to the analytical prediction in Eq. \eqref{eq:saddlezetacal}.}
		\label{plot:alphadependencemassivecorrectionchargedmoments}
	\end{figure}
	\begin{figure}
		\includegraphics{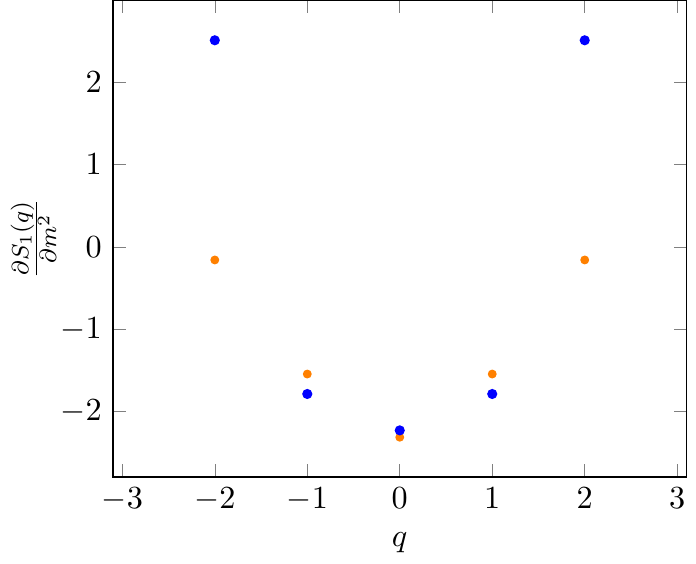}
		\includegraphics{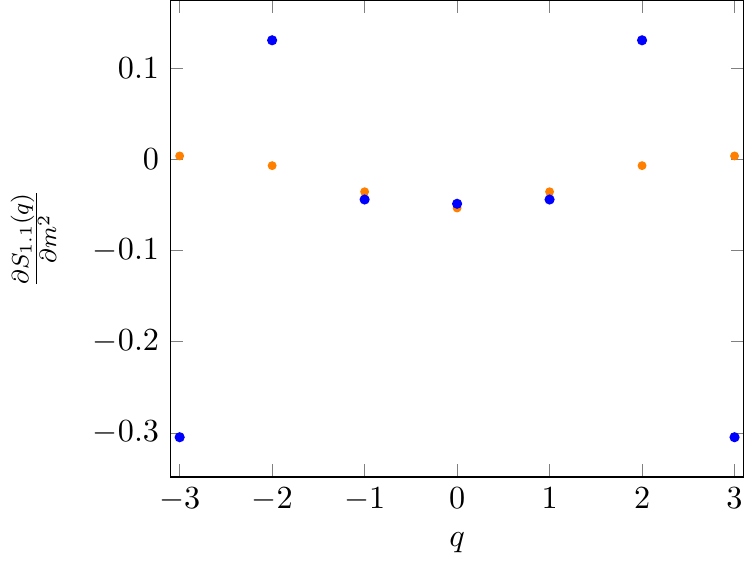}			
		\caption{Mass correction to the symmetry resolved entanglement entropy evaluated for a subsystem of size $\ell=0.5 L$ and $N=300$ total sites. The inverse temperature and the spin sector shown are $\beta=5,		 \nu=2$ (left) and $\beta=1, 		\nu=3$ (right). 
			We compare the results on the lattice (in orange) and the field theory correction (in blue) in Eq. \eqref{masscorrectionsymmresentr}.}\label{plot:SREEmassivecorr2}
	\end{figure}
	
	Similarly,  we compare the leading order mass correction of the charged moments in Fig. \ref{plot:alphadependencemassivecorrectionchargedmoments}.
	The integration of Eqs. \eqref{eq:toderive} and \eqref{eq:entropy perturbation} has been carried out with the Mathematica Montecarlo algorithm, averaged on many different realisations, to reduce the integration error, which was not negligible in a single run of the algorithm.
	\subsection{Fourier transform and number entropy}
	Starting from the charged moments in Eq. \eqref{eq:toderive}, the leading order massive correction to the symmetry resolved moments and entropy are given	by
	\begin{align}\label{eq:saddlezetacal}
		\delta \mathcal{Z}_n(q) \equiv \eva{\pdv{\mathcal{Z}^m_n(q)}{m^2}}{m^2=0}=\int_{-\pi}^\pi e^{i\alpha q} Z_n^{m=0}(\alpha)\int A^{tot}_n(\alpha,x,y)\d^2 x \d^2 y \frac{\d \alpha}{2\pi}
	\end{align} and \begin{align}
		\eva{\frac{\partial S^m_n(q)}{\partial m^2}}{m=0}=\f{1}{1-n}\left(\f{\delta \mathcal{Z}_n(q)}{\mathcal{Z}_n(q)}-n\f{\delta \mathcal{Z}_1(q)}{\Op{Z}_1(q)}\right)\label{masscorrectionsymmresentr}\end{align}
	respectively.
	The presence of a mass gives, within this  perturbative approach, a term which violates the equipartition, and which is subleading in the small $m$ limit we are considering. The correction to the symmetry resolved entropy has been plotted in Fig. \ref{plot:SREEmassivecorr2}  against the exact lattice computations; we can observe a deviation between the lattice data and the field theory results for $q>1$, which are due to finite size effect.
	
	Before concluding the section, we want to show how the mass correction to the Dirac action affects the result for the number entropy. Using a quadratic approximation for the logarithm of the charged moments around $\alpha=0$, we can get an expression for the number entropy using the saddle point approximation: by defining \begin{align}
		\int\pdv[2]{A^{tot}_{1}}{\alpha}\d^2 x \d^2 y \equiv c^\nu
	\end{align}
	we can rewrite Eq. \eqref{eq:toderive} as
	\begin{align}
		\log(Z^m_1(\alpha))\approx\log(Z_1^{m}(0))+\f{\alpha^2}{2}\left(m^2 c^\nu	- b^\nu_1\right) \end{align}
	and, using Eq. \eqref{eq:ft}, the probability distribution $p(q)$ reads
	\begin{align}
		p(q)\approx\int_{-\infty}^\infty e^{-\f{\alpha^2(b_1^\nu-m^2c^\nu)}{2}}e^{-i\alpha q}\f{\d\alpha}{2\pi} = e^{-\f{q^2}{2(b_1^\nu-m^2c_1^\nu)}}\f{1}{\sqrt{2\pi(b_1^\nu-m^2	c^\nu )}}.
	\end{align} 
	Estimating the sum with an integral, a substitution justified in the same limit of the saddle entropy approximation, the number entropy defined in Eq. \eqref{eq:sum} is
	\begin{align}
		S^{num}=-\sum_q p(q)\log(p(q)) \approx- \int_{-\infty}^\infty p(q) \log(p(q))			\d q = \unm \left(1+\log(2\pi(b_1^\nu-m^2	c^\nu )) \right),
	\end{align}
	where for small $m$, we can find the $O(m^2)$ corrections to the massless case.

	\section{Charge imbalace resolution - massive case}\label{sec:fneg}
	
	We now turn to the computation of the fermionic negativity. We need to evaluate $N_1(\alpha)$ for the normalisation constant $p(q)$ in Eq. \eqref{momentstransform} and the even charged moments 
	\begin{align}
		N_{n_e}(\alpha)=
		\Tr(\left(\rho_A^{R_1}{\rho_A^{R_1}}\dagg\right)^{n_e		} e^{i\alpha Q_{imb}}),\end{align}
	which will be analytically continued to $n_e\ra1$.
	This quantity has been studied in Ref. \cite{Murciano_2020} for massless free fermions at finite size and temperature and we will briefly review it as a starting point to compute the leading order massive correction. 
	
	As custom, we denote with $A_1$ and $A_2$ the subpartitions of $A$ and we want to evaluate the charge imbalance negativity between them.
	The odd and even charged moments $N_n(\alpha)$ can be found via replica trick; in this case we need to sew the $n$ copies of the fermionic field taking into account also the antiperiodic boundary connection to each field due to the anticommuting nature of fermions, i.e. \begin{align}
		\psi_j(x)=\begin{cases}
			T_\alpha^{R}\psi_{j-1}(x) \qquad &x\in A_1\\
			T_\alpha\psi_{j-1}(x) \qquad &x\in A_2\\
			\psi_j(x) \qquad &x\notin A
		\end{cases},\end{align}
	where $T_\alpha$ is the twist matrix already defined in Eq. \eqref{twistmatrix1}, while $T^R_\alpha$ takes into account the time-reversal operation and reads
	\begin{align}
		T^{R_1}_\alpha=\begin{pmatrix}
			0&0&0&\ldots&0	&e^{-i\frac{\alpha}{n_e}} (-1)^{n_e-1 }\\
			e^{-i\frac{\alpha}{n_e}}&0&0&\ldots&0	&0\\
			0&e^{-i\frac{\alpha}{n_e}}&0&\ldots&0	&0\\
			0&0&e^{-i\frac{\alpha}{n_e}}&\ldots&0	&0\\
			\vdots&\vdots&\vdots&\vdots&\vdots&\vdots\\
			0&0&0&\ldots&e^{-i\f{\alpha}{n}}&0	
		\end{pmatrix}.\label{twistmatrix2}
	\end{align}
	The twist matrices $T_{\alpha}$ and $T^{R_1}_{\alpha}$, defined in Eqs. \eqref{twistmatrix1} and \eqref{twistmatrix2}, respectively, can be simultaneously diagonalised for $n$ even or, trivially, for $n=1$, by choosing the following basis for the fields on the replicas
	\begin{align}
		\tilde{\psi}_k\equiv \sum_{j=1}^n e^{-\frac{2\pi i kj}{n_e}-i\pi j} \psi_j, \qquad 
		T_{\alpha}\tilde{\psi}_k=e^{\frac{i(2\pi  k+\alpha)}{n_e}}\tilde{\psi}_k\qquad T^R_{\alpha}\tilde{\psi}_k=e^{\frac{-i(2\pi  k+\alpha)}{n_e}+i\pi}\tilde{\psi}_k,\label{diagonalization_twistfield}\end{align}
	for $k=\frac{1-n_e}{2},\ldots\frac{n_e-1}{2}$.
	In this basis, our problem is split into the computation of $n$ decoupled partition functions in which the fields have different twist phases,  $e^{2\pi i \frac{k+\alpha/2\pi}{n_e}}$ and $-e^{-2\pi i \frac{k+\alpha/2\pi}{n_e}}$ passing through $A_2$ and $A_1$, respectively. The total partition function is the product of each of these
	\begin{align}
		N_{n_e}(\alpha)=\prod_k Z_{R_1,k}(\alpha).
	\end{align}
	As a concrete example, we consider the following geometry: our subsystems $A_1,A_2$ are adjacent segments of lengths $\ell_1,\ell_2$, respectively, such that $A_1=[-{\ell_1},0], 
	A_2=[0,\ell_2]$. 
	Moreover, we work at finite size and temperature, so that each partition function is defined on a torus.
	Using again the bosonisation technique, $ Z_{R_1,k}(\alpha)$ is expressed as a correlation function of vertex operators. As explained in Appendix \ref{fluxambiguity} we need to fix the phase ambiguity of the vertex operators  by choosing the leading order contribution in the scaling limit. This has been done in Appendix \ref{ambiguity_negativity}, and we report here only the final result
	\begin{align}
		Z_{R_1,k}=\ave{e^{- i \left(\f{\sgn\left(k\right)}{2}+\f{k}{n_e}+\f{\alpha}{2\pi n_e}\right)\phi(-\ell_1)}e^{ i \left(\f{\sgn(k)}{2}+\f{2k}{n_e}+\f{\alpha}{\pi n_e}\right)\phi(0)}e^{ i \left(\f{k}{n_e}+\f{\alpha}{2\pi n_e}\right)\phi(\ell_2)}}\label{eq31},
	\end{align} 
	which is readily evaluated for massless fermions as \cite{Murciano_2020}
	\begin{align}
		\begin{split}
			Z^{m=0}_{R_1,k}(\alpha)=&\f{\abs{\theta(-\ell_1+\ell_2|\tau)}^{2\left(\f{k}{n_e}-\f{\sgn{(k)}}{2}+\f{\alpha}{2\pi n_e}\right)\left(\f{k}{n_e}+\f{\alpha}{2\pi n_e}\right)}}{\abs{\theta_1(\ell_1|\tau)}^{2\left(\f{k}{n_e}-\f{\sgn{(k)}}{2}+\f{\alpha}{2\pi n_e}\right)\left(\f{2k}{n_e}-\f{\sgn{(k)}}{2}+\f{2\alpha}{n_e}\right)}\abs{\theta_1(\ell_2|\tau)}^{2\left(\f{k}{n_e}+\f{\alpha}{2\pi n_e}\right)\left(\f{2k}{n_e}-\f{\sgn{(k)}}{2}+\f{2\alpha}{n_e}\right)}}\times\\&\times
			\abs{\f{\theta_\nu\left[\left(\ell_2-\ell_1\right)\f{k+\alpha/2\pi}{n_e}+\ell_1\f{\sgn(k)}{2}\right]}{\theta_\nu(0|\tau)}}^2\abs{\epsilon\theta'(0|\tau)}^{\Delta_{k,n}},\\
			\Delta_{k,n}=&\left(\f{\sgn\left(k\right)}{2}+\f{k}{n_e}+\f{\alpha}{2\pi n_e}\right)^2+\left(\f{\sgn(k)}{2}+\f{2k}{n_e}+\f{\alpha}{\pi n_e}\right)^2+\left(\f{k}{n_e}+\f{\alpha}{2\pi n_e}\right)^2.
		\end{split}
	\end{align}	
	Here we want to study the leading order massive correction using again a perturbative expansion, which mirrors the one in Section \ref{Section:massive_calculations}. 
	This expansion is meaningful only if the mass is taken to be smaller than all the energy scales involved, i.e. $m\ll T, L^{-1}$.	 
	In this case we define the quantity $\tilde{A}_{k,n}$ as
	\begin{align}\tilde{A}_{k,n}=
		\f{\ave{e^{i\left(\phi(x)-\phi(y)\right)}\left(e^{i \left(\f{k+\alpha/2\pi}{n_e}-\f{\sgn(k)}{(k+\alpha/2\pi)}\right)\phi(-\ell_1)}e^{-i \left(\f{2(k+\alpha/2\pi)}{n_e}-\f{\sgn(k)}{2}\right)\phi(0)}e^{i \f{k+\alpha/2\pi}{n_e}\phi(\ell_2)}\right)}}{\ave{e^{i \left(\f{k+\alpha/2\pi}{n_e}-\f{\sgn(k)}{2}\right)\phi(-\ell_1)}e^{-i 
					\left(\f{2(k+\alpha/2\pi)}{n_e}-\f{\sgn(k)}{2}\right)\phi(0)}e^{i \f{k+\alpha/2\pi}{n_e}\phi(\ell_2)}}},
	\end{align}
	which is again a correlator of vertex operator that can be explicitly worked out.
	Thus, the massive correction can be written as
	\begin{align}
		\log(Z_{R_1,k}^m)\approx\log(Z_{R_1,k}^{m=0})+m^2 \int  \tilde{A}_{k,n}(\alpha,x,y)-\eva{\tilde{A}_{k,n}(\alpha,x,y)}{k=\alpha=0} \d^2 x\d^2 y,
		\label{eq:espansionefunzionepartneg}
	\end{align}
	\begin{align}
		\tilde{A}_{k,n}=\abs{\left(\f{\theta_1(x+\ell_1)\theta_1(x-\ell_2)\theta_1(y)^2}{\theta_1(y+\ell_1)\theta_1(y-\ell_2)\theta_1(x)^2}\right)^\f{k+\alpha/2\pi}{n_e}\left[
			\f{\theta_1(x+\ell_1)\theta_1(y)}{\theta_1(y+\ell_1)\theta_1(x)}\right]^{-\f{\sgn(k)}{2}}}^2\times\nonumber\\
		\times \abs{\f{\epsilon \theta'_1(0|\tau)\theta_\nu(\f{k+\alpha/2\pi}{n_e}(\ell_2-\ell_1)+\ell_1\f{\sgn(k)}{2}+x-y)}{\theta_1(x-y|\tau)\theta_\nu(\f{k+\alpha/2\pi}{n_e}(\ell_2-\ell_1)+\ell_1\f{\sgn(k)}{2})}}^2.
		\label{eq:10}	\end{align}
	The integral in Eq. \eqref{eq:espansionefunzionepartneg}, similarly to the one in Eq. \eqref{kdependent massive correction to the partition function}, does not have any UV divergences.
	This expression can be analytically continued to
	\begin{align}
		\tilde{A}^{tot}_n(\alpha,x,y)=\sum_k  \tilde{A}_{k,n}(\alpha,x,y)-\eva{\tilde{A}_{k,n}(\alpha,x,y)}{k=\alpha=0},
	\end{align}
	for all values of $n_e\ge 1$. In  Appendix \ref{app:resummation_negativity}, we report its explicit expression, which has been used to obtain the numerical plots.
	\begin{figure}			\includegraphics{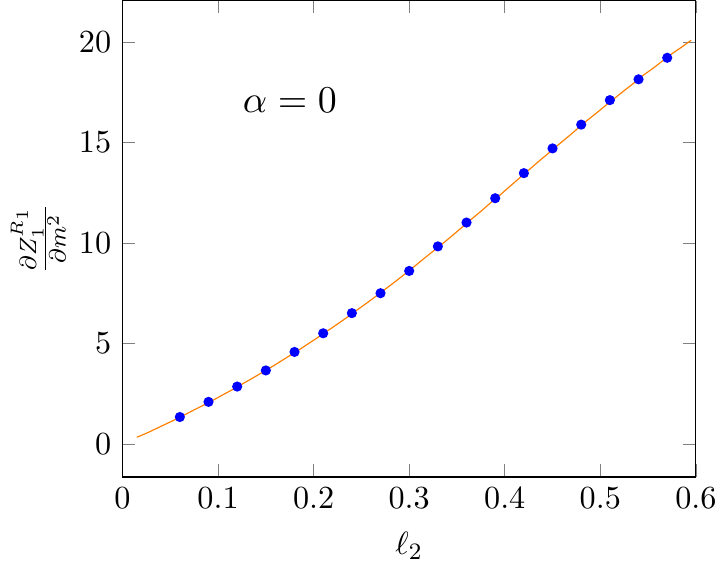}
		\includegraphics{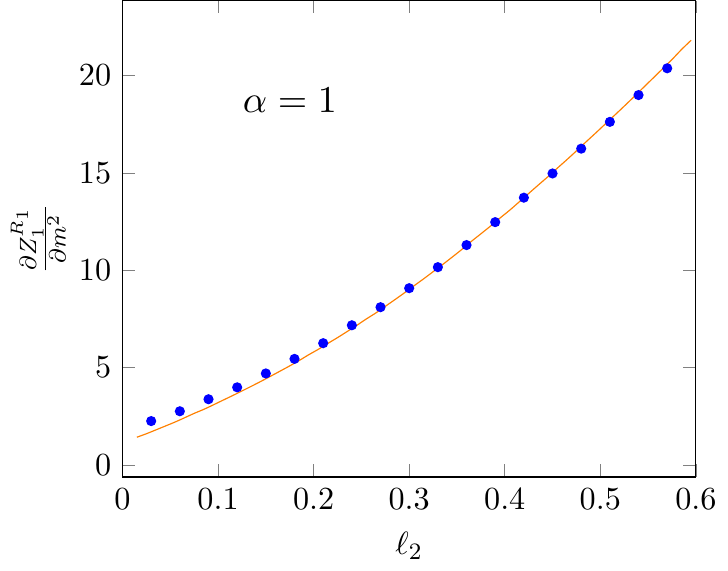}						
		\includegraphics{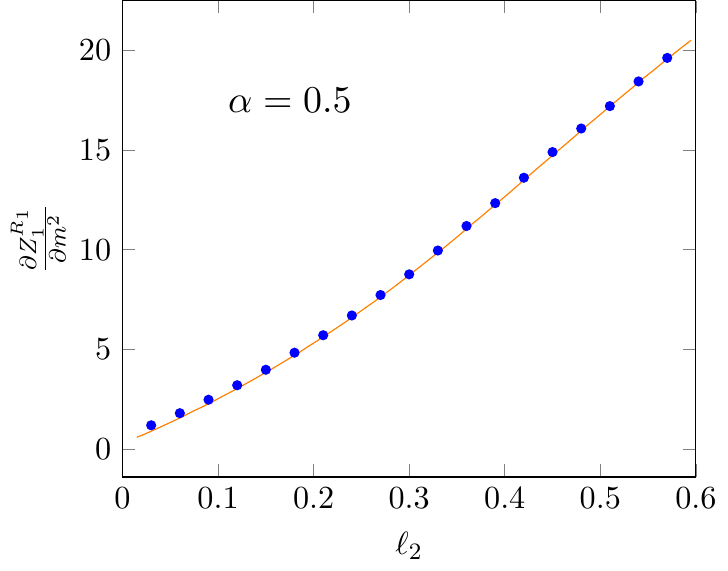}
		\includegraphics{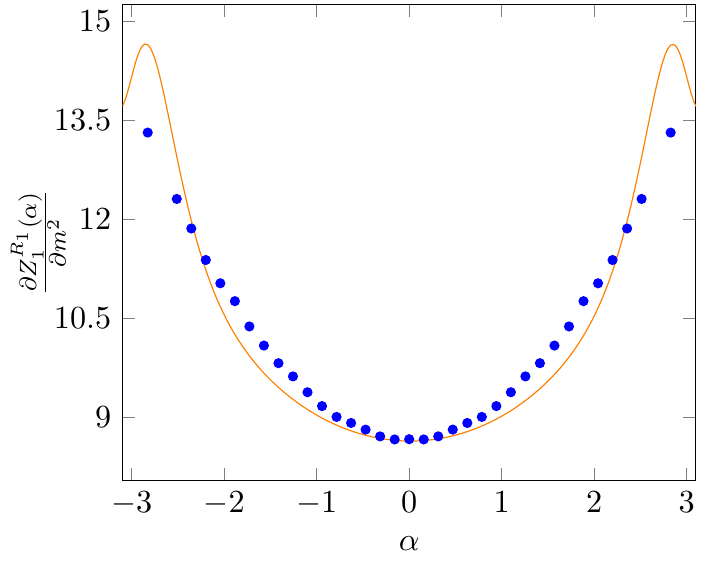}		
			\label{plot:SREEmassivecorr}
		\caption{Top and bottom left: massive correction to the partial transposed moment $Z_{n_e}^{R_1}(\alpha)$, for $n_e=1$, $\beta=10$, as a function of the length of the interval $A_2$, $\ell_2$. $A_1$ and $A_2$ are two adjacent intervals, where the size of $A_1$ is fixed and equal to $0.4$ (we chose the total spacial length of the system to be $L=1$ with $N=300$ sites ). The value of $\alpha$ is fixed and shown in the legend. In blue we plot the field theory prediction, in orange the lattice data. Bottom right: $\alpha$-dependence of the massive correction for the partial transposed moment. The subsystems are adjacent and their size is fixed to be $\ell_1=0.4 $ and $\ell_2=0.3$. The blue points are the field theory prediction in Eq. \eqref{eq:negtoplot}, while the solid orange line describes the lattice data. }
		\label{plot:dependencemassivecorrectionNeg}
	\end{figure}
	The case $n=1$, where $n$ is considered odd and not an analytic continuation from the even values, can be treated explicitly:	\begin{align}
I(\alpha,x,y)\equiv	\f{\ave{e^{i\left(\phi(x)-\phi(y)\right)}e^{ -i \left(\Phi_\alpha+\f{\alpha}{2\pi }\right)\phi(-\ell_1)}e^{ i \left(\Phi_\alpha+\f{\alpha}{\pi }\right)\phi(0)}e^{ -i \left(\Phi_\alpha+\f{\alpha}{2\pi }\right)\phi(\ell_2)}}}{\ave{e^{ -i \left(\left(\Phi_\alpha+\f{\alpha}{2\pi }\right)\phi(-\ell_1)\right)}e^{ i \left(\Phi_\alpha+\f{\alpha}{\pi }\right)\phi(0)}e^{ -i \left(\Phi_\alpha+\f{\alpha}{2\pi }\right)\phi(\ell_2)}}},\nonumber\\
		\delta(({Z}_{R_1,n=1}^m))=m^2\int \d^2 x\d^2 y \left(I(\alpha,x,y)-I(0,x,y)\right)\:,
	\end{align}
	where the phase ambiguity is fixed by the prescription\begin{align}
		\Phi_\alpha=\begin{cases}
			0 &\abs{\alpha}\le\f{2\pi}{3},\\
			-\sgn{(\alpha)}  &\abs{\alpha}\ge\f{2\pi}{3},
		\end{cases}
	\end{align}as derived in Appendix \ref{ambiguity_negativity}, and the correlation function is evaluated using Eq. \eqref{eq:corrfunctiontorus}.
	
	In Figure \ref{plot:dependencemassivecorrectionNeg}, we plot the analytic continuation of \begin{align}\label{eq:negtoplot}
		\f{1}{Z_{R_1,n_e=1}(\alpha)}\eva{\pdv{{Z}_{R_1,n_e=1}(\alpha)^m}{m^2}}{m}\int \tilde{A}^{tot}_n \d^2 x \d^2 y.
	\end{align}
	In the first three plots, we fix the value of $\alpha$ to $\alpha=0,0.5,1$, the size of $A_1$ is $0.4 L$ and the size of $A_2$ corresponds to the $x$ axis; in the last panel, instead, we have fixed the sizes of $A_1, A_2$  and we let  the value of $\alpha$ vary in the range $[-\pi,\pi]$. In every case, we compare the field theory result with the ones obtained from the lattice, fitting a mass renormalisation constant.
	Moreover, the evaluation of the integral in Eq. \eqref{eq:negtoplot}, performed with the Mathematica Montecarlo algorithm, is particularly challenging given the oscillations of the integrand, which may lead to small, systematic errors.
	We also made the assumption, for the cutoff constant $\epsilon$ in Eq. \eqref{eq:10}, to be independent on $\alpha$, which allows it to be absorbed in the definition of $\lambda$, and that may explain a small mismatch between the lattice theory and the field theory in the last plot of Figure \ref{plot:dependencemassivecorrectionNeg}. 
	Finally, following the steps outlined in Section \ref{sec:details_neg}, one can link the charged quantities to the charge-imbalance negativity.
	The massive correction to the Fourier transforms can be expressed as
	\begin{align}\label{eq:saddlezetacalneg}
		\delta \mathcal{Z}_{R_1,n}(q) \equiv \eva{\pdv{\mathcal{Z}^m_{R_1,n}(q)}{m^2}}{m^2=0}=\int_{-\pi}^\pi e^{-i\alpha q} Z_{R_1,n}^{m=0}(\alpha)\int \tilde{A}^{tot}_n(\alpha,x,y)\d^2 x \d^2 y \frac{\d \alpha}{2\pi},
	\end{align} 
	while the charge-imbalance resolved negativities read
	\begin{align}
		\eva{\frac{\partial \mathcal{N}(q)}{\partial m^2}}{m=0}=\frac{1}{2}\left(\f{\delta \mathcal{Z}_{R_1,n_e=1}(q)}{\mathcal{Z}_{R_1,n_e=1}(q)}-\f{\delta \mathcal{Z}_{R_1,n=1}(q)}{\Op{Z}_{R_1,n=1}(q)}\right).\label{masscorrectionsymmresneg}\end{align}
	
	Since we are working in the regime in which the mass correction is smaller than the other inverse lengths, the correction is subleading with respect to the  results found in \cite{Murciano_2021} for massless fermions, so we find again equipartition at leading order. 	
	
	\section{Conclusions}\label{sec:conclusion}
	
	In this manuscript, we have studied the behaviour of entanglement in systems with a conserved local charge explicitly computing both the symmetry resolved entropies and the charge imbalance resolved negativity. 
	We first considered the 1+1 dimensional CFT of free Dirac fermions at finite temperature $T$ and finite size $L$,  distinguishing the boundary conditions imposed on the fermionic field in the spacial and temporal directions. 
	We focused on the sectors with antiperiodic conditions in the imaginary time and periodic ($\nu=2$) or antiperiodic ($\nu=3$) along the space. The other two spin sectors (i.e. $\nu=1,4$) have not been considered here, since they correspond to periodic conditions in the imaginary time, while physical fermions are required to be antiperiodic. 
	We found that while the spin-independent part of the symmetry resolved entropies satisfies entanglement equipartition, the spin dependent term, i.e. the one related to the boundary conditions, causes
	a small subleading violation of equipartition.
	The comparisons between the analytical formulae and the exact lattice computations showed an overall very good agreement. 
	
	We then moved to the massive Dirac fermions treating the mass as a perturbation of the CFT. We found that the correction due to the small mass violates equipartition, showing that, when moving away from criticality, systems tend to part from this behavior. This result generalises the findings of Ref. \cite{Murciano_2020}, in the planar limit, to the torus. 
	As a byproduct, we have also managed to analytically continue the expression of the leading massive correction to the total R\'enyi entropies.
	We also studied how the presence of the mass affects the charge imbalance resolved-negativity, always in conformal perturbation theory. Also for the negativity, the massive term breaks the equipartition.
	
	There are some aspects that our manuscript leaves open for further investigations. For example, the sum over all the possible spin sectors on the torus would lead to the symmetry resolved entanglement for the modular invariant Dirac fermion. Furthermore, our formalism strictly relies on the decoupling of the fermionic modes in the replica space, which is true only for free theories. It is natural to wonder what happens for interacting fermion (i.e. a compact boson with given compactification radius). To this aim, we could employ some of the methods developed in the literature as, e.g., in Ref. \cite{multicharged}.
	
	\section*{Acknowledgements}
	P.C. acknowledges support from ERC under Consolidator grant number 771536 (NEMO). SM is supported by Walter Burke Institute for Theoretical Physics at Caltech.
	
	\begin{appendix}

		\section{Numerical methods}\label{app:num}
		
		In this first appendix, we report how to numerically calculate the charged moments of the reduced density matrix $\rho_A$. 
		In free lattice models, $\rho_A$ can be written as \cite{Peschel_2009,Peschel_2003,Chung_2001} \begin{equation}
			\rho_A=\f{1}{Z}e^{-\Op{H}},\label{effective Hamiltonian definition}
		\end{equation}
		where $\Op{H}$ is a free effective entanglement (or modular) Hamiltonian. 
		Using the relation between the entanglement Hamiltonian and the correlation matrix restricted to the subsystem $A$, $\vec{C}$, the charged moments on the lattice 
		$Z_n(\alpha)=\Tr {\rho^n_Ae^{ iQ_A\alpha}}$ are \cite{Goldstein_2018, Bonsignori_2019}
		\begin{equation}
			Z_n(\alpha)=\log(\Tr{\vec{C}^n e^{2 \pi i \alpha \vec{N}}+\left(\vec{\unity-C}\right)^n})
		\end{equation}
		where the charge (or number) operator $\Op{N}$ distinguishes between particles and antiparticles
		\[\Op{N}=\sum_l f\dagg_l f_l-g\dagg_l g_l,\numberthis \]
		with $g\dagg_l$ the creation operator for antiparticles, and $f\dagg_l$ for particles.
		
		The entanglement Hamiltonian is also the sum of two parts which contain particles and antiparticles, respectively:
		\[\Op{H}=\sum_l \epsilon_l f\dagg_lf_l+\epsilon_l g\dagg_l g_l=\Op{H}_p+\Op{H}_a. \numberthis \]
		The correlation matrix is similarly divided in two blocks 
		\[\vec{C}=\begin{pmatrix}\vec{P}&0\\0&\vec{A}
		\end{pmatrix} \numberthis \]
		since $\ave{g\dagg f}=0$, and $\vec{P},\vec{A}$ are the particles/antiparticles restricted correlation matrices, respectively.
		Restricting to particles, $\vec{N}=\unity$, so that
		\[\log(Z^p_n(\alpha))=\Tr{\log(e^{2\pi  i \alpha\unity}\vec{P}^n+\left({1-\vec{P}}\right)^n)}, \label{symmetryresolved expressionparticles}\numberthis \]
		while for antiparticles, $\vec{N}=-\unity$ and
		\[\log(Z^a_n(\alpha))=\Tr{\log(e^{-2\pi  i \alpha\unity}\vec{A}^n+\left({1-\vec{A}}\right)^n)}. \numberthis \]
		Assuming we have the same number of states for particles and antiparticles (i.e. the total state is neutral):
		\begin{multline}
			\log(Z_n(\alpha))=\left[\Tr{\log(e^{-2\pi  i \alpha\unity}\vec{A}^n+\left({\unity-\vec{A}}\right)^n)}+\Tr{\log(e^{2\pi  i \alpha\unity}\vec{P}^n+\left({\unity-\vec{P}}\right)^n)}\right]=\\
			\Tr{\log(e^{-i\pi{\alpha}{}\unity}\vec{A}^n+e^{i\pi {\alpha}{}\unity}\left({\unity-\vec{A}}\right)^n)}+\Tr{\log(e^{i\pi{\alpha}{}\unity}\vec{P}^n+e^{-i\pi{\alpha}{}\unity}\left({\unity-\vec{P}}\right)^n)}.\label{latticefermionrenyientropy}
		\end{multline}
		
		We now need a proper lattice discretisation of the field theory. The free (massive) fermion Lagrangian in Minkowsky space is
		\[\Op{L}=\int  \bar{\psi}\left(i\slashed{\partial}-m	\right)\psi\d x^1, \label{Minkowsky lagrangian}\numberthis \]
		where $\slashed{\partial}=\partial_\mu \gamma^\mu$, $\bar{\psi}=\psi^{\dagger} \gamma^0$ and the matrices $\gamma^\mu$ obey the Clifford algebra
		\[\{\gamma^\mu,\gamma^\nu\}=2\eta^{\mu\nu}\unity \qquad \eta^{\mu\nu}=\mathrm{diag}\begin{pmatrix}+&-\end{pmatrix}.\numberthis \]
		A representation of such matrices in 2D is
		\[\gamma^0=\sigma^1=\begin{pmatrix}
			0&1\\
			1&0
		\end{pmatrix}\qquad\gamma^1=-i\sigma^2=\begin{pmatrix}
			0&-1\\
			1&0
		\end{pmatrix}. \numberthis \]
		If the total length of the lattice is $L$ (in particular we set $L=1$) and we have $N$ sites, the discretisation rules are:
		\[\epsilon\equiv \f{L}{N}\qquad \partial_1\longrightarrow\f{1}{\epsilon}\Delta\qquad \int \d x^1\longrightarrow\epsilon \sum_j \qquad\delta(x)\longrightarrow \delta_{jk}\f{1}{\epsilon}. \numberthis \]
		It is also useful to rescale the fields, obtaining the canonical anticommutation relations
		\[\{\psi(x),\psi(y)\}=\delta(x-y)\ra \{\Psi_i,\Psi_j\}=\f{\delta_{ij}}{\epsilon}\implies \{\psi_i,\psi_j\}={\delta_{ij}}\qquad \psi_i\equiv\Psi_i \sqrt{\epsilon}. \numberthis \]
		Then, the discretised Lagrangian is
		\[\f{1}{\epsilon}\sum_j {\psi}_j\dagg\left(i\partial_0\epsilon-m\epsilon\sigma^1	\right)\psi_j+\unm\psi_j\dagg i\sigma^3\psi_{j+1}-\unm\psi_{j+1}\dagg i\sigma^3\psi_j. \numberthis \]
		In order to get the Hamiltonian, we perform a Legendre transformation, which yields 
		\[ H=\f{1}{\epsilon}\sum_j^N\unm\left(\psi_{j+1}\dagg i\sigma^3\psi_j-\psi_j\dagg i\sigma^3\psi_{j+1}\right)+m \epsilon \psi\dagg_{j} \sigma^1 \psi_j.\label{discretized Dirac Hamiltonian}\numberthis \]
		Being the Hamiltonian quadratic, it can be explicitly diagonalised. 
		In particular, we can notice it is invariant under translations $j\ra j+1$, so the eigenvectors are of the form (with $\nu_1=0,\unm$ corresponding to periodic or antiperiodic spacial fermions, respectively, and with Greek letters for spinor indices):
		\[f^\alpha_{k,n}=\sum_{j,\beta} \f{1}{\sqrt{N}}e^{\f{2\pi i (k-\nu_1)j}{N}}\tensor{\left(\Op{M}^{-1}\right)}{^\alpha^\beta}\psi_j^\beta \numberthis \]
		This relation can be inverted 
		\begin{equation}
			{\psi_j}^\alpha=\sum_{\beta\in\{1,2\},k=1}^{N}\f{1}{\sqrt{ N}}\tensor{\Op{M}}{^\alpha^\beta} e^{-\f{2\pi i (k-\nu_1)j}{N}}f^\beta_{k,n} .\label{field operators diagonalized}\end{equation}
		Plugging it into the Hamiltonian \eqref{discretized Dirac Hamiltonian}, the diagonalisation reads
		\begin{equation}
			\tan(\theta)\equiv\f{m\epsilon}{\sin(\f{2\pi (k-\nu_1)}{N})}\implies\left(\Op{M}\right)^{\pm 1}=\cos(\f{\theta}{2})\unity\pm i\sin(\f{\theta}{2})\sigma_2\label{matrixchangeofbasis}\end{equation}
		The energy eigenvalues are \cite{Herzog_2013} \[\epsilon_{\pm,k}=\pm\sqrt{m^2+\f{\sin(\f{2\pi (k-\nu_1)}{N})^2}{\epsilon^2}}\label{eigenvalues} \numberthis \]
		where the negative eigenvalues physically represent the annihilation of antiparticles.
		As noted in \cite{Herzog_2013}, in Eq. \eqref{eigenvalues}, the eigenvalues $\epsilon_{\pm,k}$ are equal to the $\epsilon_{\pm,N-k}$ (this is true for $\nu_1=0$, in the antiperiodic case $\nu_1=1/2$ the correspondence is $k\leftrightarrow N-k+1$) and so a doubling problem is present. 
		We simply fix the doubling problem by dividing the final result by two.
		
		The correlator of a Dirac field includes both particles and antiparticles:
		\[\ave{f\dagg_k f_q}=\f{\delta_{kq}}{1+e^{\epsilon_k}}, 
		\qquad\ave{g_k g\dagg_q}=\delta_{kq}-\ave{g\dagg_q g_k}=\delta_{kq}-\f{\delta_{kq}}{1+e^{\epsilon_k}},\qquad \ave{g_k f_q}=\ave{g\dagg_k f\dagg_q}=0, \numberthis \]
		and the implicit form of the correlation matrix of the Dirac field is		\begin{multline}\label{finalformlatticecorr}
			\vec{C}_{(l,\alpha),(m,\beta)}\equiv\ave{{\psi_l^\alpha}\dagg{\psi_m^\beta}}=\\
			=\sum_k \f{e^{\f{2\pi i (k-\nu_1)(l-m)}{N}}}{{N}}\left[ {(\Op{M}^{\alpha+})}^*(k)\Op{M}^{\beta+}(k)\ave{f_k\dagg f_k }+\ave{g_k g\dagg_k} {(\Op{M}^{\alpha-})}^*(k)\Op{M}^{\beta-}(k)\right].
		\end{multline}
		This shows that the spinor indices in the correlation matrix $\psi\dagg\psi$ contain the particle correlation function $P_{lm}$, and the matrix $\ave{g_m g\dagg_l}=\left(\unity-\vec{A}\right)_{ml}$.
		Using  Eq. \eqref{latticefermionrenyientropy}, we get
		\begin{equation}
			Z_n(\alpha)=\unm \left[\Tr{\log(e^{i\pi{\alpha}{}\unity}\vec{C}^n+e^{-i\pi{\alpha}{}\unity}\left({\unity-\vec{C}}\right)^n)}\right],
			\label{fermionlatticemoments}
		\end{equation}
		where the $\unm$ factor has been inserted to cure of the doubling problem.
		Using Eq. \eqref{finalformlatticecorr}, we can now compute $\vec{C}$ explicitly. The correlators in the eigenstate basis are \[\ave{{f^{\alpha}_l}\dagg f^{\beta}_m}=\unm\left(\unity+\sigma_3\tanh(\f{\epsilon_k}{2})\right)\qquad f^+_k=f_k\qquad f^-_k=g\dagg_k	 \numberthis \]
		By plugging the explicit form of $\Op{M}$ \eqref{matrixchangeofbasis} into Eq. \eqref{finalformlatticecorr}, we get
		\begin{equation}
			C_{jh}=\sum_k \f{e^{\f{2\pi i (k-\nu_1)(j-h)}{N}}}{2N}\left(\unity+\f{\sin(\f{2\pi i (k-\nu_1)}{N})\sigma_3+m\epsilon\sigma_1}{\sqrt{\sin(\f{2\pi i (k-\nu_1)}{N})^2+(m\epsilon)^2}}\tanh(\f{\beta\epsilon_k}{2})\right).
			\label{correlation matrix, massive case}
		\end{equation}
		In the massless case we have instead \cite{Herzog_2013}
		\begin{equation}
			C_{jh}=\sum_k \f{e^{\f{2\pi i (k-\nu_1)(j-h)}{N}}}{2N}\left(\unity+{\text{sgn}\left(\sin(\f{2\pi i (k-\nu_1)}{N})\right)\sigma_3}\tanh(\f{\beta\epsilon_k}{2})\right).
			\label{correlation matrix, massless case}
		\end{equation}
		Moreover, there is an additional subtlety in the choice of $N$: as noted in \cite{Herzog_2013}, for $N$ odd we end up with two copies of the system (which are due to the doubling problem), of opposite spacial periodicity. In order to only have one fixed boundary condition, the simplest solution is to limit ourselves   to take $N$ even.
		
		\subsection{Fermionic negativity}\label{sec:numerics_negativity}
		We want now to compute the entanglement negativity for the same system of lattice free fermions.
		We switch to the Majorana fermion operators $c_{i,\alpha}$:
		\begin{align}
			c_{2i-1,\alpha}=f_{i,\alpha}+f_{i,\alpha}\dagg\qquad			c_{2i,\alpha}=i\left(f_{i,\alpha}-f_{i,\alpha}\dagg\right)
		\end{align}
		Any operator can be expressed as a sum of string of Majoranas, in the form
		\begin{align}
			\rho=\sum_{k  \text{ even}}\sum_{\{a_i\}_{i=1}^k}\rho_{\{a_i\}} c_{a_1}c_{a_2}\ldots c_{a_2}.
		\end{align}
		We took the string length to be even, which is a constraint for theories conserving the parity fermion number operator.	
		Following \cite{Murciano_2021}, we consider the covariance matrix $\Gamma$, that can be expressed as $\sigma_2\otimes (\unity-2 C)\equiv \sigma_2\otimes \Gamma$, with $C$ being the correlation matrix \eqref{correlation matrix, massive case}. Our total system is bipartite into $A,B$, and we want to compute the entanglement between two parts of $A$, which are $A_1, A_2$. The reduced covariance matrix on $A$ takes the block form \begin{align}
			\Gamma=\begin{pmatrix}
				\Gamma_{AA} & \Gamma_{AB}\\
				\Gamma_{BA} & \Gamma_{BB}\\
			\end{pmatrix}.
		\end{align} 
		Under partial transposition, we find the correlation matrices associated to $\rho^{R_1}$, ${\rho^{R_1}}\dagg$, which we call $\Gamma_{\pm}$\begin{align}
			\Gamma_\pm=\begin{pmatrix}
				-\Gamma_{AA} & \pm i\Gamma_{AB}\\
				\pm i\Gamma_{BA} & \Gamma_{BB}\\
			\end{pmatrix}.
		\end{align}	
		We also introduce
		\begin{equation}\label{eq:gammax}
			\Gamma_{\mathrm{x}}=(\mathbbm{1}+\Gamma_+\Gamma_-)^{-1}(\Gamma_++\Gamma_-).
		\end{equation}
		Finally, with the eigenvalues of $\Gamma_{\mathrm{x}}$ and $C$, which we call $\nu_i$ and $\zeta_i$, respectively we can express the even charged moments of the partial transposed density matrix, as\begin{multline}
			\log(N_{n_e})=\\-i\alpha\f{\ell_1+\ell_2	}{2}+\sum_j\log\left[\left(\f{1-\nu_j}{2}\right)^\f{n}{2}+\left(\f{1+\nu_j}{2}\right)^\f{n}{2} e^{i\alpha}\right]+\f{n}{2}\sum_j \log\left[\zeta^2+(1-\zeta_j)^2\right].  
		\end{multline}		
		We also need the charged normalisation $N_1$, which can be written in terms of the eigenvalues of $\Gamma_+$, $\nu^+_j$, as \cite{Murciano_2021}
		\begin{align}
			\log(N_1)=-i\alpha\f{\ell_1+\ell_2}{2}+\sum_j\log\left[\left(\f{1-\nu^+_j}{2}\right)+e^{i\alpha}\left(\f{1+\nu^+_j}{2}\right)\right].
		\end{align}

		\section{Flux strength ambiguity}\label{fluxambiguity}
		As observed in Ref. \cite{Shapourian_2019}, Eq. \eqref{Gaugefield definition} is actually ambiguous, since we can add an integer number of $2\pi$ flux caused by the gauge field $A_\mu$, such that
		\[\f{k+\alpha/2\pi}{n}\ra\f{k+\alpha/2\pi}{n}+m, \qquad m\in \mathbb{N}, \numberthis \]
		without affecting the monodromy of the fermion fields.
		
		The partition function $Z_k$ must preserve this symmetry, so the correct form of the partition function is a combination of all the possible \emph{inequivalent} representations each corresponding to a different value of $m$.
		Without losing generality, we suppose to compute the correlation functions of $p$ vertex operators with phases $h_i$, respecting the neutrality condition\begin{align}
			\sum_i h_i=0.
		\end{align}
		We call $m_i$ the integer addition to the flux, which must respect the neutrality $\sum_i m_i=0$, too.
		The most general partition function is a combination of all the choices of $\{m_i\}_{i=1,\ldots,p}$\begin{equation}
			\begin{split}
				Z=&\sum_{\{m_i\}} C_{\{m_i\}}\ave{\prod_i e^{\left(h_i+m_i\right)\phi(x_i)}}=\\=&\sum_{\{m_i\}} C_{\{m_i\}} \abs{\f{\abs{\theta_{\nu}(\sum_{i}\left(h_i+m_i\right)x_i|\tau)}}{\theta_\nu(0|\tau)}}^2\abs{\prod_{i<j}\f{\epsilon\partial_x \eva{\theta_1(x|\tau)}{x=0}}{\theta_1(x_i-x_j)}}^{-2 (h_i+m_i)(h_j+m_j)}.
			\end{split}	\label{robba}	\end{equation}
		In the end, we want to test our field theoretical predictions against the numerical results for the theory defined on a lattice.
		This implies that the cutoff behaves as $\epsilon\propto a$, where $a=\frac{L}{N}$ is the lattice spacing, given by the ratio between the total spacial lenght of the torus, which we set $L=1$ in Appendix \ref{app:num}, and the number of lattice sites. This can be seen by either a dimensional argument, or by noting that, if we compute on the lattice the correlation function of two vertex operators, we find \begin{align}
			\ave{e^{i\alpha \phi(0)}e^{-i\alpha \phi(a)}}\sim 1
		\end{align}
		because the two opposite vertex operators almost overlap on the lattice.
		Field theory would predict, instead
		\begin{align}
			\ave{e^{i\alpha \phi(0)}e^{-i\alpha \phi(a)}}=\abs{\f{\theta_\nu(\alpha a|\tau)}{\theta_\nu(0|\tau)}}^2\abs{\f{\epsilon \theta'_1(0|\tau)}{ \theta_1(a|\tau)}}^{2\alpha^2}=\abs{\f{\epsilon \theta'_1(0|\tau)}{ a\theta_1'(0|\tau)}+o(a^2)}^{2\alpha^2}
			, \end{align}
		where we have done an expansion at the lowest order in $a$. Comparing the two expressions we get $\epsilon\sim a$.
		Going back to Eq. \eqref{robba}, we focus on the $\epsilon$ factor, which is:
		\begin{align}
			\prod_{i<j}\epsilon^{-2(h_i+m_i)(h_j+m_j)}=\left(\f{\prod_{i,j}\epsilon^{-2(h_i+m_i)(h_j+m_j)}}{\prod_{i}\epsilon^{-2(h_i+m_i)^2}}\right)^\unm=\epsilon^{(h_i+m_i)^2},
		\end{align}
		where we used the neutrality condition $\prod_i \epsilon^{-2(h_i+m_i)}=1$.
		In the limit $\epsilon\sim a\ra0$  the leading term in \eqref{robba} will be the one minimising the exponent of $\epsilon$, which is $\sum_i(h_i+m_i)^2$.
		\subsection{Entanglement entropy}
		In the entanglement entropy calculations, rewriting the partition function in Eq. \eqref{partitionfunctiondefinition} as
		\begin{equation}
			Z_k=\ave{\prod_j\exp( i\left(\f{k+\alpha/(2\pi)}{n}+{m}_a \right)\phi(u_j,0))\exp( i\left(-\f{k+\alpha/(2\pi)}{n}+\bar{m}_a \right)\phi(v_j,0))},\label{rewritten partition function definition}\end{equation}
		the neutrality condition of vertex operators imposes
		\begin{equation}\sum_a m_a+\bar{m}_a=0.\label{neutrality condition extra fluxes}\end{equation}
		This factor is maximised by the following choices:
		\begin{equation}
			\begin{split}\label{alphacondition}
				\bar{m_i}=-m_i,\quad 
				\Big |\f{k+{\alpha}/({2\pi})}{n}+{m}_a \Big |<\unm.
			\end{split}    
		\end{equation}
	Since $n\ge 1$, $|k|\le \f{n-1}{2}, |\alpha|\le \pi,$
	we have that $
	|\f{k+\alpha/(2\pi)}{n}|<\unm,$
	thus all the dominant $m_a$'s have to be $0$.
	This justifies retrospectively our calculations, and also predicts a worse agreement with the lattice computations for $|\alpha|\approx \pi$, where the power law suppression with $\epsilon^{2\left(\f{k+\alpha/2\pi}{n}\right)^2}$ is much weaker and contested by a $m=-1$ flux, for example with a term of the form $\sim \epsilon^{2\left(\f{k+\alpha/2\pi}{n}-1\right)^2}$. Thus, when plotting the $\alpha$ dependence of the charged moments, we expect a worse agreement at the edges of the figure (as can be well seen in Figs. \ref{diffnu2}-\ref{plot:alphadependencechargedmoments}). Similar conclusions were also drawn in \cite{Bonsignori_2019}, where the subleading corrections were found to be comparable to the dominant ones around the edges $\alpha=\pm \pi$. 
	\subsection{Fermionic negativity}\label{ambiguity_negativity}
	For the negativity in the tripartite geometry, the partition function has, in general, the form 
	\begin{align}
		Z_{R_1,k}=\ave{e^{- i \left(\unm+m_1+\f{k}{n_e}+\f{\alpha}{2\pi n_e}\right)\phi(-\ell_1)}e^{ i \left(\unm+m_1+m_2+\f{2k}{n_e}+\f{\alpha}{\pi n_e}\right)\phi(0)}e^{- i \left(m_2+\f{k}{n_e}+\f{\alpha}{2\pi n_e}\right)\phi(\ell_2)}},
	\end{align}		
	for $n=n_e$ even. As before, we need to minimise the sum\begin{align}
		\left(m_1+\unm +\f{k}{n_e}+\f{\alpha}{2\pi n_e}\right)^2+	\left(\f{2k}{n_e}+\unm+\f{\alpha}{\pi n_e}+m_1+m_2\right)^2+	\left(m_2+\f{k}{n_e}
		+\f{\alpha}{2\pi n_e}\right)^2,
	\end{align}
	where we already imposed the neutrality condition for the extra fluxes $m_a$. 
	When $k+\f{\alpha}{2\pi}\ge0$ we need to require $m_1=-1$, i.e. $m_2=1$, while, in	 the remaining case, $m_1=0$ ($m_2=0$).
	The lowest positive value of $k$ we can take is $k=\f{n_e-1}{2}=\f{1}{2}\ge\f{\alpha}{2\pi}$, since $\alpha\in[-\pi,\pi]$. This implies $\sgn(k+\f{\alpha}{2\pi})=\sgn(k)$, leading to the correct expression:\begin{align}
		Z_{R_1,k}\sim\ave{e^{- i \left(-\f{\sgn\left({k}\right)}{2}+\f{k}{n_e}+\f{\alpha}{2\pi n_e}\right)\phi(-\ell_1)}e^{ i \left(\f{\sgn\left({k}\right)}{2}+\f{2k}{n_e}+\f{\alpha}{\pi n_e}\right)\phi(0)}e^{- i \left(\f{k}{n_e}+\f{\alpha}{2\pi n_e}\right)\phi(\ell_2)}}.\label{flux-corrected_correlator}
	\end{align} 
	
	Additionally, in the $n=1$ case, we need to consider
	\begin{align}
		Z_{R_1,n=1}=\ave{e^{- i \left(m_1+\f{\alpha}{2\pi }\right)\phi(-\ell_1)}e^{ i \left(m_1+m_2+\f{\alpha}{\pi }\right)\phi(0)}e^{- i \left(m_2+\f{\alpha}{2\pi }\right)\phi(\ell_2)}}.
	\end{align}
	The quantity 
	\begin{align}
		\left(m_1+\f{\alpha}{2\pi }\right)^2+	\left(\f{\alpha}{\pi }+m_1+m_2\right)^2+	\left(m_2
		+\f{\alpha}{2\pi }\right)^2,
	\end{align}
	is minimised, if $\abs{\alpha}\le \f{2\pi}{3}$, for the value $m_1=m_2=0$, while, in the remaining cases, for either $(m_1,m_2)=(-\sgn(\alpha),0),(0,-\sgn(\alpha))$.
	
	\section{Saddle point approximation in the massless case}\label{app:saddlepoint}
	In the limit of the cutoff $\epsilon$ going to  $0$, the leading order for the alpha dependence of the logarithm of the charged moments is \begin{align}
		\log(Z_n(\alpha))\approx \log(Z_n(\alpha=0))+\frac{\alpha^2}{2\pi^2 n}\log(\epsilon)+O(\epsilon^0).
	\end{align}
	This allows us to perform an expansion around $\alpha^2=0$ in order to obtain the leading order contribution from the other terms, as has been done in Eq. \eqref{eq:saddlepoint expansion}. The coefficients are obtained by expanding Eqs. \eqref{eq:chm}, \eqref{eq:nu2spindeppart}, \eqref{eq:nu3spindeppart}:
	\begin{align}
		\begin{split}
			b^{\nu=2}_n\equiv &-\f{1}{ \pi^2 n}\log\left(\f{\prod_{a<b}\theta_1(\f{u_a-u_b}{i\beta}|-\f{1}{\tau}=\f{i}{\beta})\theta_1(\f{v_a-v_b}{i\beta}|-\f{1}{\tau}=\f{i}{\beta})}{\prod_{a,b}\theta_1(\f{u_a-v_b}{i\beta}|-\f{1}{\tau}=\f{i}{\beta})}\left(\f{ \epsilon\partial_z \theta_1(0|\tau)}{ i\beta }\right)^p\right)+\\&+2\sum_{m \ge 1}	\left(\f{  r}{ n\beta}\right)^2\f{m}{2\sinh(\f{\pi m L}{\beta} )}\left({\f{\sinh{\f{\pi r m}{\beta}}}{\sinh{\f{\pi r m	}{\beta n}}}}\right)+p\frac{1-n^2}{6n\epsilon}\eva{\pdv[2]{\epsilon(n,\alpha)}{\alpha}}{\alpha=0},\label{divergnu3}\\
			b^{\nu=3}_n\equiv& -\f{1}{\pi^2 n}\log\left(\f{\prod_{a<b}\theta_1(\f{u_a-u_b}{i\beta}|-\f{1}{\tau}=\f{i}{\beta})\theta_1(\f{v_a-v_b}{i\beta}|-\f{1}{\tau}=\f{i}{\beta})}{\prod_{a,b}\theta_1(\f{u_a-v_b}{i\beta}|-\f{1}{\tau}=\f{i}{\beta})}\left(\f{ \epsilon\partial_z \theta_1(0|\tau)}{ i\beta }\right)^p\right)+\\&+2\sum_{m \ge 1}	(-1)^m \left(\f{r}{ n\beta}\right)^2\f{m}{\sinh(\f{\pi m L}{\beta} )}\left({\f{\sinh{\f{\pi r m}{\beta}}}{\sinh{\f{\pi r m	}{\beta n}}}}\right)+p\frac{1-n^2}{6n\epsilon}\eva{\pdv[2]{\epsilon(n,\alpha)}{\alpha}}{\alpha=0}, \\
			A_n=&{\Big|\f{\prod_{j<k}\theta_1(u_j-u_k|\tau)\theta_1(v_j-v_k|\tau)}{\prod_{j,k}\theta_1(u_j-v_k|\tau)}(\epsilon \partial_z \theta_1(0|\tau))^p\Big|}^\f{n^2-1}{6n}\\
			&\times\exp{2\sum_{m=1}^\infty \f{1}{m\sinh(\f{\pi m L }{\beta})} \left(n-\f{\sinh(\f{\pi m r}{\beta })}{\sinh(\f{\pi m r}{\beta n})}\right)}.
		\end{split}
	\end{align}

	\section{Analytical continuation of the symmetry resolved entanglement entropy }
	\label{analytical_sum_ssEE}		Here we report the details for the analytical continuation of the symmetry resolved entanglement in the massive fermionic theory for any real values of $n$. This is the main step to take the replica limit $n \to 1$. 
	\subsection{$\nu=2$ spin sector}
	A modular transformation of Eq. \eqref{kdependent massive correction to the partition function} leads to
	\begin{multline}		A_{k,n}=\left|\f{\theta_{\nu'}\left(\f{(k+\alpha/2\pi) r}{i\beta n}+\f{x-y}{i\beta}|-\f{1}{\tau}\right)}{\theta_{\nu'}\left(\f{(k+\alpha/2\pi) r}{i\beta n}|-\f{1}{\tau}\right)}\f{\epsilon\partial_z\theta_1\left(0|-\f{1}{\tau}\right)}{\beta \theta_1\left(\f{x-y}{i\beta}|-\f{1}{\tau}\right)}\right|^2\\
		\times
		\prod_a\left|\f{\theta_1\left(\f{v_a-x}{i\beta}|-\f{1}{\tau}\right)\theta_1\left(\f{u_a-y}{i\beta}|-\f{1}{\tau}\right)}{\theta_1\left(\f{v_a-y}{i\beta}|-\f{1}{\tau}\right)\theta_1\left(\f{u_a-x}{i\beta}|-\f{1}{\tau}\right)}\right|^\f{2(k+\alpha/2\pi)}{n},\label{modular transformed mass correction}	\end{multline}	
	with $\nu'=4$, since $\nu=2$.
	The resummation can be carried out rewriting $A_{k,n}$ with the following identity, \cite{whittaker_watson_1996}
	\begin{equation}\label{whittakerformula}
		\f{\theta_1(x+y|\tau)\theta_1'(0|\tau)}{\theta_1(x|\tau)\theta_1(y|\tau)}=\pi(\cot(y\pi)+\cot(x\pi)+4\sum_{m,n=0}^\infty e^{2mni\pi \tau}\sin(2\pi(mx+ny))),
	\end{equation}
	combined with the quasiperiodicity relation 
	\begin{equation}\jacobitheta{4}{z}{\tau}=i e^{-i\pi z} e^{\f{i\pi\tau}{4}}\jacobitheta{1}{z-\f{\tau}{2}}{\tau}.
	\end{equation}
	$A_{k,n}$ is rewritten as
	\begin{equation} \begin{split} \sqrt{ A_{k,n}} &= \f{\epsilon}{\beta}e^{-\pi \f{ \Re{x-y}	}{\beta}}e^{\f{\Phi(k+\alpha/(2\pi) )}{n}}\left|\coth({\f{x-y}{\beta}}{\pi})+\coth(\f{\pi}{\beta}{}\left(\f{(k+\alpha/(2\pi) ) r}{n}+\unm\right))+\right.\nonumber\\ &\left.-4 \sum_{m,l=1}^{\infty}  e^{-\f{2ml \pi}{\beta}}\sinh(\f{2\pi}{\beta } \left(m (x-y)+ l\left(\f{(k+\alpha/(2\pi) ) r}{n}+\unm\right) \right))\right|. \label{nu2 expression to sum} \end{split} \end{equation}
	and 
	\begin{equation}
		\Phi(x,y) \equiv\log\left(\prod_a|\f{\theta_1\left(v_a-x|\tau\right)\theta_1\left(u_a-y|\tau\right)}{\theta_1\left(v_a-y|\tau\right)\theta_1\left(u_a-x|\tau\right)}|\right).	\label{def:phifunction}
	\end{equation}
	
	When Eq. \eqref{nu2 expression to sum} is squared, we get $6$ terms that we can index schematically as  $K_i$  \[\sum_k A_{k,n}= \f{\epsilon^2}{\beta^2}e^{-2\pi \f{ \Re{x-y}	}{\beta}}\sum_{k=-\f{n-1}{2}}^{\f{n-1}{2}}\left(|K_1|^2+|K_2|^2+|K_3|^2+2 \Re{K_1^*K_2+K_2^*K_3+K_1^*K_3}\right), \numberthis\]
	where $i\in\{1,2,3\}$:
	\begin{align}
		K_1=&e^{\f{\Phi(k+\alpha/(2\pi) )}{n}} \coth({\f{x-y}{\beta}}{\pi}),\label{def:K1}\\
		K_2=&e^{\f{\Phi(k+\alpha/(2\pi) )}{n}} \coth(\f{\pi}{\beta}{}\left(\f{(k+\alpha/(2\pi) ) r}{n}+\unm\right)),\label{def:K2}\\			K_3=&-4e^{\f{\Phi(k+\alpha/(2\pi) )}{n}} \sum_{m,l=1}^{\infty} e^{-\f{2ml \pi}{\beta}}\sinh(\f{2\pi}{\beta } \left(m (x-y)+ l\left(\f{(k+\alpha/(2\pi) ) r}{n}+\unm\right) \right)).\label{def:K3}
	\end{align}
	We sum in $k$ each addendum in this equation separately, labelling the  results as $\gamma_{ij}$:
	\[\gamma_{ii}\equiv\sum_{k=-\f{n-1}{2}}^{\f{n-1}{2}}|K_i|^2\qquad\gamma_{ij}\equiv2\sum_{k=-\f{n-1}{2}}^{\f{n-1}{2}} \Re{K_iK_j^*}\qquad j\ne i\qquad i,j=1,2,3.\label{Eq:sumoverk} \numberthis\]
	Finally, we follow the prescription in Eq. \eqref{eq:rego}, which amounts to change the $\gamma$s as\begin{align}
		\gamma^{reg}=\gamma_{ij}-\gamma_{ij}(\alpha=0, r=0),
	\end{align}
	so that the final result is written as\begin{align}
		A^{tot}_n=\f{\epsilon^2}{\beta^2}e^{-2\pi \f{ \Re{x-y}	}{\beta}}\sum_{i,j}\gamma^{reg}_{ij}.
	\end{align}
	The analytic continuation in $n$  of expressions \eqref{Eq:sumoverk} can be found since every term in Eqs. \eqref{def:K1}, \eqref{def:K2}, \eqref{def:K3} can be expressed as a convergent sum of exponentials, for the range of parameters we are interested in.
	In particular, we can rewrite the $\coth$ function with the identity:\begin{align}
		\coth(x)=\text{sign}(x)\left(1+2\sum_{m=1}^{\infty} e^{-2 m\abs{x}}\right).
		\label{coth expansion}
	\end{align}
	Using this fact, we can reduce the sum  over $k$ in \eqref{Eq:sumoverk} to a geometric series with a trivial analytic continuation. 
	We now report the final expressions found using this approach:
	\[\gamma_{11}=\f{\sinh({\Phi })}{\sinh(\f{\Phi}{n})} e^{\f{\Phi\alpha }{\pi n}}\abs{ \coth(\f{x-y}{\beta}\pi)}^2 \numberthis \]
	\[\gamma_{12}=2 \Re{\coth(\f{x-y}{\beta}\pi)}e^{\f{\Phi \alpha }{\pi n}}\left(\f{\sinh(\Phi)}{\sinh(\f{\Phi}{n})}+2\sum_{m=1}^\infty e^{-\f{2\pi m}{\beta}\left(\f{\alpha r}{2\pi n}+\unm\right)}\f{\sinh(\Phi-\f{\pi m r}{\beta})}{\sinh(\f{1}{n}\left({\Phi-\f{\pi m r}{\beta}}\right))}\right).\numberthis \]
	\[\gamma_{22}=e^{\f{\Phi \alpha }{\pi n}}\left(\f{\sinh{\Phi }}{\sinh(\f{\Phi}{n})}+4\sum_{m=1}^\infty m e^{-\f{2\pi m}{\beta}\left(\f{\alpha r}{2\pi n}+\unm\right)}\f{\sinh(\Phi-\f{\pi m r}{\beta})}{\sinh(\f{1}{n}\left({\Phi-\f{\pi m r}{\beta}}\right))}\right). \numberthis \]
	
	\[\gamma_{33}={\gamma_{33}^{++}+\gamma_{33}^{--}}-2 \Re{\gamma_{33}^{-+}}. \numberthis \]
	\[\gamma_{13}=-4 \Re{\coth(\f{\bar{x}-\bar{y}}{\beta})\Sigma_{13}\exp(\f{\alpha \Phi}{\pi n})}, \numberthis \]
	\[\gamma_{23}=-4\left(\Sigma_{13}+2 \Sigma_{23}\right)\exp(\f{\alpha \Phi}{\pi n}), \numberthis \]
	We used the following auxiliary expressions:
		\begin{align}
		\gamma_{33}^{++}= 4 e^{\f{\Phi \alpha }{\pi n}} \sum_{l,q=1}^{\infty} \exp(\f{2\pi}{\beta } (l+q){}\left(\unm+\f{\alpha r}{2\pi n}\right))\f{\sinh(\Phi+\f{\pi(l+q)r}{\beta})}{\sinh({\f{\Phi}{n}+\f{\pi(l+q)r}{n\beta}})}\times\nonumber\\\times\f{1}{\exp(\f{2\pi}{\beta}(l-x+y))-1}\f{1}{\exp(\f{2\pi}{\beta}(q-\bar{x}+\bar{y}))-1}, \end{align}
	
	\begin{align}
		\gamma_{33}^{--}= 4 e^{\f{\Phi \alpha }{\pi n}}\sum_{l,q=1}^{\infty} \exp(-\f{2\pi}{\beta } (l+q){}\left(\unm+\f{\alpha r}{2\pi n}\right))\f{\sinh(\Phi-\f{2\pi(l+q)r}{\beta})}{\sinh({\f{\Phi}{n}-\f{2\pi(l+q)r}{n\beta}})}\times\nonumber\\\times\f{1}{\exp(\f{2\pi}{\beta}(l+x-y))-1}\f{1}{\exp(\f{2\pi}{\beta}(q+\bar{x}-\bar{y}))-1},
	\end{align}
	\begin{align}
		\gamma_{33}^{-+}=4 e^{\f{\Phi \alpha }{\pi n}}\sum_{l,q=1}^{\infty} \exp(-\f{2\pi}{\beta } (l-q){}\left(\unm+\f{\alpha r}{2\pi n}\right))\f{\sinh(\Phi-\f{2\pi(l-q)r}{\beta})}{\sinh({\f{\Phi}{n}-\f{2\pi(l-q)r}{n\beta}})}\times\nonumber\\\times\f{1}{\exp(\f{2\pi}{\beta}(l+x-y))-1}\f{1}{\exp(\f{2\pi}{\beta}(q-\bar{x}+\bar{y}))-1} ,
	\end{align}
	\begin{equation}\label{eq::sigma13}\begin{gathered}
			\Sigma_{13}=
			\sum_{l=1}^{\infty}\left( e^{\f{\pi \left(2r\alpha/(2\pi) +n\right) l}{\beta n}}\f{\sinh(\Phi+\f{\pi l r}{\beta})}{\sinh(\f{\Phi}{n}+\f{\pi l r }{\beta n})}\f{1}{\exp(\f{2\pi}{\beta}(l-x+y))-1}+\right.\\\left.-e^{-\f{\pi \left(2r\alpha/(2\pi) +n\right) l}{\beta n}}\f{\sinh(\Phi-\f{\pi l r}{\beta})}{\sinh(\f{\Phi}{n}-\f{\pi l r }{\beta n})}\f{1}{\exp(\f{2\pi}{\beta}(l+x-y))-1}\right),\end{gathered}\end{equation}	
	\begin{align}\begin{gathered}
			\Sigma_{23}=
			\sum_{l,p=1}^{\infty} e^{\f{\pi \left(2r\alpha/(2\pi) +n\right) (l-p)}{\beta n}}\f{\sinh(\Phi+\f{\pi (l-p) r}{\beta})}{\sinh(\f{\Phi}{n}+\f{\pi (l-p) r }{\beta n})}\f{1}{\exp(\f{2\pi}{\beta}(l-x+y))-1}\\-e^{-\f{\pi \left(2r\alpha/(2\pi) +n\right) (l+p)}{\beta n}}\f{\sinh(\Phi-\f{\pi (l+p) r}{\beta})}{\sinh(\f{\Phi}{n}-\f{\pi (l+p) r }{\beta n})}\f{1}{\exp(\f{2\pi}{\beta}(l+x-y))-1}.\end{gathered}
	\end{align}

	\subsection{$\nu=3$ spin sector}\label{nu=3 massive case resummation}
	Similar calculations can be carried out in this spin sector $\nu=3$. 
	We can write again:
	\begin{align}
		A^{tot}_n=\frac{\epsilon^2}{\beta^2} e^{-\frac{2\pi \Re{x-y}}{\beta}}
		\sum_{i,j}\gamma^{reg}_{ij}.	\end{align}
	We use again expression \eqref{def:phifunction} for $\Phi$, while we need to modify those for $\gamma_{ij}$ in this spin sector, as following:\[\gamma_{11}=\f{\sinh({\Phi })}{\sinh(\f{\Phi}{n})} e^{\f{\Phi\alpha }{\pi n}}\abs{ \coth(\f{x-y}{\beta}\pi)}^2, \numberthis \]
	\begin{equation}        
		\gamma_{12}=2 \Re{\coth(\f{x-y}{\beta}\pi)}e^{\f{\Phi \alpha }{\pi n}}\left(\f{\sinh(\Phi)}{\sinh(\f{\Phi}{n})}+2\sum_{m=1}^\infty(-)^m e^{-\f{2\pi m}{\beta}\left(\f{\alpha r}{2\pi n}+\unm\right)}\f{\sinh(\Phi-\f{\pi m r}{\beta})}{\sinh(\f{1}{n}\left({\Phi-\f{\pi m r}{\beta}}\right))}\right),\numberthis 
	\end{equation}
	\[\gamma_{22}=e^{\f{\Phi \alpha }{\pi n}}\left(\f{\sinh{\Phi }}{\sinh(\f{\Phi}{n})}+4\sum_{m=1}^\infty(-)^m m e^{-\f{2\pi m}{\beta}\left(\f{\alpha r}{2\pi n}+\unm\right)}\f{\sinh(\Phi-\f{\pi m r}{\beta})}{\sinh(\f{1}{n}\left({\Phi-\f{\pi m r}{\beta}}\right))}\right) ,\numberthis \]
	\[\gamma_{13}=-4 \Re{\coth(\f{\bar{x}-\bar{y}}{\beta})\Sigma_{13}\exp(\f{\alpha \Phi}{\pi n})}, \numberthis \]
	
	\[\gamma_{23}=-4\left(\Sigma_{13}+2 \Sigma_{23}\right)\exp(\f{\alpha \Phi}{\pi n}), \numberthis \]
	\[\gamma_{33}={\gamma_{33}^{++}+\gamma_{33}^{--}}-2 \Re{\gamma_{33}^{-+}} .\numberthis \]
	The auxiliary expressions used are 
	\begin{align}
		\gamma_{33}^{++}= 4 e^{\f{\Phi \alpha }{\pi n}} \sum_{l,q=1}^{\infty} (-)^{l+q}\exp(\f{2\pi}{\beta } (l+q){}\left(\unm+\f{\alpha r}{2\pi n}\right))\times\nonumber\\\times\f{\sinh(\Phi+\f{\pi(l+q)r}{\beta})}{\sinh({\f{\Phi}{n}+\f{\pi(l+q)r}{n\beta}})}\f{1}{\exp(\f{2\pi}{\beta}(l-x+y))-1}\f{1}{\exp(\f{2\pi}{\beta}(q-\bar{x}+\bar{y}))-1}, \end{align}
	
	\begin{align}
		\gamma_{33}^{--}= 4 e^{\f{\Phi \alpha }{\pi n}}\sum_{l,q=1}^{\infty} (-)^{l+q}\exp(-\f{2\pi}{\beta } (l+q){}\left(\unm+\f{\alpha r}{2\pi n}\right))\times\nonumber\\\times\f{\sinh(\Phi-\f{2\pi(l+q)r}{\beta})}{\sinh({\f{\Phi}{n}-\f{2\pi(l+q)r}{n\beta}})}\f{1}{\exp(\f{2\pi}{\beta}(l+x-y))-1}\f{1}{\exp(\f{2\pi}{\beta}(q+\bar{x}-\bar{y}))-1},
	\end{align}
	\begin{align}
		\gamma_{33}^{-+}=4 e^{\f{\Phi \alpha }{\pi n}}\sum_{l,q=1}^{\infty} (-)^{l+q}\exp(-\f{2\pi}{\beta } (l-q){}\left(\unm+\f{\alpha r}{2\pi n}\right))\times\nonumber\\\times\f{\sinh(\Phi-\f{2\pi(l-q)r}{\beta})}{\sinh({\f{\Phi}{n}-\f{2\pi(l-q)r}{n\beta}})}\f{1}{\exp(\f{2\pi}{\beta}(l+x-y))-1}\f{1}{\exp(\f{2\pi}{\beta}(q-\bar{x}+\bar{y}))-1},
	\end{align}
	\begin{align}\begin{gathered}
			\Sigma_{13}=
			\sum_{l=1}^{\infty} (-)^l \left(e^{\f{\pi \left(2r\alpha +n\right) l}{\beta n}}\f{\sinh(\Phi+\f{\pi l r}{\beta})}{\sinh(\f{\Phi}{n}+\f{\pi l r }{\beta n})}\f{1}{\exp(\f{2\pi}{\beta}(l-x+y))-1}\right.+\nonumber\\\left.-e^{-\f{\pi \left(2r\alpha +n\right) l}{\beta n}}\f{\sinh(\Phi-\f{\pi l r}{\beta})}{\sinh(\f{\Phi}{n}-\f{\pi l r }{\beta n})}\f{1}{\exp(\f{2\pi}{\beta}(l+x-y))-1}\right)\end{gathered},\end{align}
	
	\begin{align}\begin{gathered}
			\Sigma_{23}=
			\sum_{l,p=1}^{\infty} (-)^{l+p}\left[e^{\f{\pi \left(2r\alpha +n\right) (l-p)}{\beta n}}\f{\sinh(\Phi+\f{\pi (l-p) r}{\beta})}{\sinh(\f{\Phi}{n}+\f{\pi (l-p) r }{\beta n})}\f{1}{\exp(\f{2\pi}{\beta}(l-x+y))-1}+\right.\nonumber\\\left.-e^{-\f{\pi \left(2r\alpha +n\right) (l+p)}{\beta n}}\f{\sinh(\Phi-\f{\pi (l+p) r}{\beta})}{\sinh(\f{\Phi}{n}-\f{\pi (l+p) r }{\beta n})}\f{1}{\exp(\f{2\pi}{\beta}(l+x-y))-1}\right]\end{gathered}.\end{align}
\end{appendix}
\section{Analytical continuation of moments for the charge imbalance negativity }\label{app:resummation_negativity}
Here we give explicit analytical continuations for $A_k$ in Eq. \eqref{eq:10}, in the spin sectors of interest, $\nu=2$. Similar results can be obtained also for $\nu=3$, as done for the entanglement entropy.
The techniques used for the resummation are similar to the ones described in Appendix \ref{analytical_sum_ssEE}, the main difference being the fact we need to divide the case $k>0$ from $k<0$, to handle the sign function; we only report the final expressions for simplicity, after some premises.

The expression to resum, after a modular transformation, is:
\begin{align}
	\tilde{A}_{k,n}=e^{2\Phi(\f{k+\alpha/(2\pi)}{n_e})-{\Psi}{\sgn(k)}}\abs{\f{\epsilon}{i\beta}\f{\theta'_1\left(0\left|-\f{1}{\tau}\right.\right)\jacobitheta{\nu'}{\f{k+\alpha/2\pi}{i\beta n_e}(\ell_2-\ell_1)+\f{\ell_1\sgn(k)}{2i\beta}+\f{x-y}{i\beta}}{-\f{1}{\tau}}}{\jacobitheta{1}{\f{x-y}{i\beta}}{-\f{1}{\tau}}\jacobitheta{\nu'}{\f{k+\alpha/2\pi}{i\beta n_e}(\ell_2-\ell_1)+\f{\ell_1\sgn(k)}{2i\beta}}{-\f{1}{\tau}}}}^2,
\end{align}
with \begin{align}
	\Phi=&\log \left|\left(\f{\jacobitheta{1}{\f{x+\ell_1}{i\beta}}{-\f{1}{\tau}}\jacobitheta{1}{\f{x-\ell_2}{i\beta}}{-\f{1}{\tau}}\jacobitheta{1}{\f{y}{i\beta}}{-\f{1}{\tau}}}{\jacobitheta{1}{\f{y+\ell_1}{i\beta}}{-\f{1}{\tau}}\jacobitheta{1}{\f{y-\ell_2}{i\beta}}{-\f{1}{\tau}}\jacobitheta{1}{\f{x}{i\beta}}{-\f{1}{\tau}}}\right)\right|,\\
	\Psi=&\log \left|
	\f{\jacobitheta{1}{\f{x+\ell_1}{i\beta}}{-\f{1}{\tau}}\jacobitheta{1}{\f{y}{i\beta}}{-\f{1}{\tau}}}
	{\jacobitheta{1}{\f{y+\ell_1}{i\beta}}{-\f{1}{\tau}}\jacobitheta{1}{\f{x}{i\beta}}{-\f{1}{\tau}}}\right|.
\end{align}
We set $\nu=2\implies \nu'=4$.
Using property \ref{whittakerformula} and definining $r=\ell_2-\ell_1$,  it becomes:
\begin{align}
	\tilde{A}_{k,n}	=&\left(\f{\epsilon}{\beta}\right)^2 e^{2\Phi(\f{k+\alpha/(2\pi)}{n_e})-{\Psi}{\sgn(k)}}e^{-2\pi\f{\Re{x-y}}{\beta}}\left|\coth(\pi\left(\f{(k+\alpha/2\pi) r}{\beta n}+\f{1+\sgn{(k)}\ell_1}{2\beta}\right))+\right.\nonumber\\&+\coth(\pi\f{x+y}{\beta})
	\nonumber\\&\left.
	-4\sum_{c,d=1}^\infty e^{-\f{2cd\pi}{\beta}}\sinh(2\pi\left( c \left(r\f{k+\alpha/2\pi}{\beta n}+\f{1+\ell_1\sgn(k)}{2\beta}\right)+d\f{x-y}{\beta}\right))\right|^2.
\end{align}
Note that this series is well defined and convergent for all values of the parameters we are interesed in, i.e. $\alpha\in[-\pi,\pi], k\in[-\frac{n-1}{2n},\frac{n-1}{2n}]$.\\	Moreover, if we assume a cutoff independent of $n$ and $\alpha$, then we can reabsorb $\epsilon$ in the definition of the mass, so we will ignore it from now on.
We define auxiliary quantities similar to what has been done in Appendix \ref{analytical_sum_ssEE}:
\begin{align}
	&\abs{K_1+K_2+K_3}^2\equiv
	\\&=e^{2\Phi\left(\f{k+\alpha/(2\pi)}{n_e}\right)-{\Psi}{\sgn(k)}}\left|\coth(\pi\f{x+y}{\beta})+\coth(\pi\left(\f{(k+\alpha/2\pi) r}{\beta n}+\f{1+\sgn{(k)}\ell_1}{2\beta}\right))+\right.\nonumber\\&\left.-4\sum_{c,d=1}^\infty e^{-\f{2cdi\pi}{\tau}}\sinh(2\pi\left( c \left(r\f{k+\alpha/2\pi}{\beta n}+\f{1+\ell_1\sgn(k)}{2\beta}\right)+d\f{x-y}{\beta}\right))\right|^2.
\end{align}
\begin{align}
	\gamma_{ii}=
	\sum_k	\abs{K_i}^2,\qquad \gamma_{ij}=\sum_k {2} 		\Re{K_i K_j^*},\end{align}
\begin{align}
	\tilde{A}^{tot}_n=\frac{e^{-2\pi \frac{\Re{x-y}}{\beta}}}{\beta^2}\sum_{i,j} \gamma^{reg}_{ij}.
\end{align}
Again, all the terms $K_{i}K_{j}^*$ can be written as a sum of exponentials, and this allows to find an analytic continuation for the sum in $k$. The final expressions are
\[ {\gamma_{11}=\coth(\pi\f{x-y}{\beta})^2 e^{\f{2\Phi \alpha}{n_e}}\f{\sinh\left({\Phi-\Psi} \right)
		+\sinh(\Psi)}{\sinh{\f{\Phi}{n_e}}}},\numberthis\]
\begin{align}
	\gamma_{22}=e^{\f{\Phi \alpha }{\pi n}}\f{\sinh\left({\Phi-\Psi} \right)
		+\sinh(\Psi)}{\sinh{\f{\Phi}{n_e}}}+\nonumber\\+4e^{\f{\Phi \alpha }{\pi n}}\sum_{m=1}^\infty m e^{-\f{2\pi m}{\beta}\left(\f{\alpha r}{2\pi n}+\unm\right)}\f{\sinh(\Phi-\Psi-\f{\pi m \ell_2}{\beta})+\sinh(\Psi+\f{\pi m \ell_1}{\beta})}{\sinh(\f{1}{n}\left({\Phi-\f{\pi m r}{\beta}}\right))}, \numberthis \end{align}
\begin{align}
	\gamma_{12}=2 \Re{\coth(\f{x-y}{\beta}\pi)}e^{\f{\Phi \alpha }{\pi n}}\left(\f{\sinh\left({\Phi-\Psi} \right)
		+\sinh(\Psi)}{\sinh{\f{\Phi}{n_e}}}\right.+\nonumber\\
	+\left.2\sum_{m=1}^\infty e^{-\f{2\pi m}{\beta}\left(\f{\alpha r}{2\pi n}+\unm\right)}\f{\sinh(\Phi-\f{\pi m \ell_2}{\beta}-\Psi)+\sinh(\Psi+\f{\pi m \ell_1}{\beta})}{\sinh(\f{1}{n}\left({\Phi-\f{\pi m r}{\beta}}\right))}\right),
\end{align}

\[ {\gamma_{13}=-4 \Re{\coth(\f{\bar{x}-\bar{y}}{\beta})\Sigma_{13}\exp(\f{\alpha \Phi}{\pi n})}}. \numberthis \]

\[ {\gamma_{23}=-4\left(\Sigma_{13}+2 \Sigma_{23}\right)\exp(\f{\alpha \Phi}{\pi n})}. \numberthis \]
\[ {\gamma_{33}={\gamma_{33}^{++}+\gamma_{33}^{--}}-2 \Re{\gamma_{33}^{-+}}}. \numberthis \]
The auxiliary expressions read
\begin{align}
	\gamma_{33}^{--}= 4 e^{\f{\Phi \alpha }{\pi n}}\sum_{l,q=1}^{\infty} e^{-\f{2\pi}{\beta } (l+q){}\left(\unm+\f{\alpha r}{2\pi n}\right)}\f{\sinh(\Phi-\f{2\pi(l+q)\ell_2}{\beta}-\Psi)+\sinh(\Psi+\f{2\pi(l+q)\ell_1}{\beta})}{\sinh({\f{\Phi}{n}-\f{2\pi(l+q)r}{n\beta}})}\times\nonumber\\\times\f{1}{\exp(\f{2\pi}{\beta}(l+x-y))-1}\f{1}{\exp(\f{2\pi}{\beta}(q+\bar{x}-\bar{y}))-1},
\end{align}

\begin{align}
	\gamma_{33}^{++}= 4 e^{\f{\Phi \alpha }{\pi n}} \sum_{l,q=1}^{\infty} e^{\f{2\pi}{\beta } (l+q){}\left(\unm+\f{\alpha r}{2\pi n}\right)}\f{\sinh(\Phi+\f{2\pi(l+q)\ell_2}{\beta}-\Psi)+\sinh(\Psi-\f{2\pi(l+q)\ell_1}{\beta})}{\sinh({\f{\Phi}{n}+\f{2\pi(l+q)r}{n\beta}})}\times\nonumber\\\times\f{1}{\exp(\f{2\pi}{\beta}(l-x+y))-1}\f{1}{\exp(\f{2\pi}{\beta}(q-\bar{x}+\bar{y}))-1},
\end{align}

\begin{align}
	\gamma_{33}^{-+}=4 e^{\f{\Phi \alpha }{\pi n}}\sum_{l,q=1}^{\infty} e^{-\f{2\pi}{\beta } (l-q){}\left(\unm+\f{\alpha r}{2\pi n}\right)}\f{\sinh(\Phi-\f{2\pi(l-q)\ell_2}{\beta}-\Psi)+\sinh(\Psi+\f{2\pi(l-q)\ell_1}{\beta})}{\sinh({\f{\Phi}{n}-\f{2\pi(l-q)r}{n\beta}})}\times\nonumber\\\times\f{1}{\exp(\f{2\pi}{\beta}(l+x-y))-1}\f{1}{\exp(\f{2\pi}{\beta}(q-\bar{x}+\bar{y}))-1} ,
\end{align}
\begin{align}
	\Sigma_{13}=    
	\sum_{l=1}^{\infty}\left( e^{\f{\pi \left(r\alpha /\pi+n\right) l}{\beta n}}\f{\sinh(\Phi+\f{\pi   l}{\beta}\ell_2-\Psi)+\sinh(-\f{\pi l }{\beta}\ell_1+\Psi)}{\sinh(\f{\Phi}{n}+\f{\pi l r }{\beta n})}\f{1}{\exp(\f{2\pi}{\beta}(l-x+y))-1}+\right.\nonumber\\\left.-e^{-\f{\pi \left(r\alpha /\pi+n\right) l}{\beta n}}\f{\sinh(\Phi-\f{\pi   l}{\beta}\ell_2-\Psi)+\sinh(\f{\pi l }{\beta}\ell_1+\Psi)}{\sinh(\f{\Phi}{n}-\f{\pi l r }{\beta n})}\f{1}{\exp(\f{2\pi}{\beta}(l+x-y))-1}\right),
\end{align}
\begin{gather}
	\Sigma_{23}=\sum_{l,p=1}^{\infty} e^{\f{\pi \left(r\alpha/\pi +n\right) (l-p)}{\beta n}}\f{\sinh(\Phi+\f{\pi (l-p) \ell_2}{\beta}-\Psi)+\sinh(-\f{\pi (l-p) \ell_1}{\beta}+\Psi)}{\sinh(\f{\Phi}{n}+\f{\pi (l-p) r }{\beta n})}\f{1}{\exp(\f{2\pi}{\beta}(l-x+y))-1}\nonumber\\-e^{-\f{\pi \left(r\alpha/\pi+n\right) (l+p)}{\beta n}}\f{\sinh(\Phi-\f{\pi (l+p) \ell_2}{\beta}-\Psi)+\sinh(\f{\pi (l+p) \ell_1}{\beta}+\Psi)}{\sinh(\f{\Phi}{n}-\f{\pi (l+p) r }{\beta n})}\f{1}{\exp(\f{2\pi}{\beta}(l+x-y))-1}.
\end{gather}

\bibliographystyle{nb}

\bibliography{biblio}

\end{document}